\documentclass[english,twocolumn,aps]{revtex4}
\usepackage[T1]{fontenc}
\usepackage[latin9]{inputenc}
\usepackage{subfigure}
\usepackage{graphicx}
\usepackage{amssymb}

\makeatletter

\providecommand{\tabularnewline}{\\}

\makeatletter

\makeatletter

\usepackage{subfigure}

\makeatletter

\usepackage{subfigure}

\makeatletter

\usepackage{subfigure}

\makeatletter

\usepackage{esint}



\makeatother

\makeatother

\makeatother

\makeatother

\makeatother

\makeatother

\usepackage{babel}

\begin{document}

\title{Noise correlations of one-dimensional Bose mixtures in optical lattices}

\author{Anzi Hu, L. Mathey, Carl J. Williams and Charles W. Clark}

\affiliation{Joint Quantum Institute,University of Maryland and National Institute
of Standards and Technology, Gaithersburg, MD 20899}

\begin{abstract}
We study the noise correlations of one-dimensional binary Bose mixtures,
as a probe of their quantum phases. In previous work \cite{anzi},
we found a rich structure of many-body phases in such mixtures, such
as paired and counterflow superfluidity. Here we investigate the signature
of these phases in the noise correlations of the atomic cloud after
time-of-flight expansion, using both Luttinger liquid theory and the
time-evolving block decimation (TEBD) method. We find that paired
and counterflow superfluidity exhibit distinctive features in the
noise spectra. We treat both extended and inhomogeneous systems, and
our numerical work shows that the essential physics of the extended
systems is present in the trapped-atom systems of current experimental
interest. For paired and counterflow superfluid phases, we suggest
methods for extracting Luttinger parameters from noise correlation
spectroscopy. 
\end{abstract}
\maketitle

\section{\label{sec:Introduction}Introduction}

%The measurement of optical intensity correlations was first performed by Hanbury Brown and Twiss and demonstrated effect of quantum interference between different detection paths for indistinguishable particles .In
In recent years, the study of noise correlations has been established
as a way of probing ultracold atom systems \cite{folling,spielman,greiner}.
First proposed in Ref. \cite{Altman}, noise correlation spectroscopy
has been discussed as a way of measuring correlation functions of
cold atom systems, such as pairing or density order \cite{Imambekov,LM,LM_Fermion,SS,fermi_noise}.
In the experiments reported in Refs. \cite{folling,spielman,greiner},
a cold atomic gas is first held in a trap, and then released from
it by turning off the trapping potential. The noise correlations are
measured as the spatial correlations of the density in the fully expanded
atomic cloud. If atomic interactions during the expansion can be ignored,
one can use the noise correlation measurements to infer momentum space
correlations in the initial state. Such an analysis has been used
to demonstrate the phase transition between superfluid (SF) and Mott
insulator (MI) states \cite{folling,spielman}, as well as the formation
of fermionic pairs \cite{greiner}.

The experimental realization of quasi-one dimensional many-body systems
with ultra-cold atoms in optical lattices has been reported in Refs.
\cite{key-1,key-2,key-3,key-5,key-6,key-7,MI-SF(1D),BlochSpin,BlochTonks}.
Characteristic features of such systems include fluctuating and competing
orders. In contrast to higher dimensional systems which exhibit long-range
orders, 1D systems typically display only quasi-orders, that are characterized
by the algebraic decay of the correlation function of the order parameter.
In Refs. \cite{fermi_noise,LM_Fermion}, noise correlations were
shown to be an effective probe of such orders in 1D Fermi systems,
for both one- and two-component systems. Similar studies have been
done for 1D bosonic systems, either in the hard-core limit \cite{AMR}
or using Luttinger liquid (LL) theory \cite{LM}. In Ref. \cite{LM},
the signature of condensates and quasi-condensates was discussed in
detail.

Noise correlations can also be used to study the phases of binary
bosonic mixtures. In such mixtures, two additional orders beyond SF
and MI are potentially present, first studied in Ref. \cite{kuklov}.
If the inter-species interaction is attractive, bosons of different
species can form a paired superfluid (PSF) state. If the interaction
is repulsive and the system is confined in a lattice at half-filling,
the bosons can form particle-hole pairs, called {}``anti-pairs''.
Such anti-pairs can then form a counter-flow superfluid (CFSF) state.
In addition, a charge density wave order (CDW) can coexist with the
three superfluid orders, which is the defining feature of supersolidity.
Numerous examples of such order have been given in Refs.\cite{supersolid,supersolid_LM}.
In Ref. \cite{anzi} we established the phase diagram of a binary
mixture exhibiting SF, PSF, CFSF and MI orders, and we showed that
each of the superfluid orders can coexist with the CDW order.

We also showed in \cite{anzi} that because the PSF and CFSF orders
are the result of inter-species pairing, they do not provide a signature
in the momentum distributions of the individual atomic species. In
this paper, we show that noise correlation measurements provide \emph{distinctive}
signals of both the PSF and CFSF orders. Ref. \cite{menotti} shows
that noise correlations characteristic of the PSF/CFSF orders can
be observed even in a system of only four atoms. Here we calculate
the noise correlation spectra from first principles, using the time-evolving
block decimation (TEBD) method \cite{Vidal}, which is supported
by analytical calculations based on LL theory. We make appropriate
comparisons between results for homogeneous and trapped systems.

To evaluate the noise correlations, we first assume ballistic expansion
and long expansion time and define the noise correlations as the density
correlations in momentum space, \begin{eqnarray}
\mathcal{G}_{aa'}(k,k') & = & \langle n_{a,k}n_{a',k'}\rangle-\langle n_{a,k}\rangle\langle n_{a',k'}\rangle\end{eqnarray}
 where $a$, $a'$ are species indices ($a,a'=1,2$), $k$, $k'$
are momenta, and $n_{a,k}$ and $n_{a',k'}$ are the occupation operators
in momentum space. We also consider the derived quantities $C_{aa'}(q)$
and $D_{aa'}(q)$, defined as \begin{equation}
C_{aa'}(2q)=\int dk\frac{\langle n_{a}(k+q)n_{a'}(k-q)\rangle}{\langle n_{a}(k+q)\rangle\langle n_{a'}(k-q)\rangle},\label{eq:C_aa}\end{equation}
 and \begin{equation}
D_{aa'}(2q)=\int dk\frac{\langle n_{a}(k+q)n_{a'}(q-k)\rangle}{\langle n_{a}(k+q)\rangle\langle n_{a'}(q-k)\rangle}.\label{eq:D_aa}\end{equation}
 Each of these quantities can capture the main features of the noise
correlations for particular types of order and can be directly measured
in experiments \cite{spielman}. We will present all our results
first in the form of $\mathcal{G}_{aa'}(k,k')$ and then use $C_{aa'}(q)$
and $D_{aa'}(q)$ to highlight the key features.

This paper is organized as follows: in Sec. \ref{sec:The-Model},
we explain our model and the different quasi-orders present in it;
in Sec. \ref{sec:Luttinger-Liquid-Calculation}, we show our LL results
and predict the generic signature of the noise correlations for different
orders. In Sec. \ref{sec:Numerical-Calculation}, we present our numerical
calculation of noise correlations for both homogeneous and trapped
systems. In Sec. \ref{sec:Conclusion} we conclude.

\section{\label{sec:The-Model}Noise correlations and quasi-orders}

We work in an approximation in which binary Bose mixtures in optical
lattices are described by a two-component Bose-Hubbard model \cite{Bose-Hubbard,anzi}.
The Hamiltonian for $M$ atoms of each species confined to an optical
lattice with $N$ sites is given by \begin{eqnarray}
H & = & -t\sum_{a=1,2}\sum_{j=1}^{N-1}(b_{a,j}^{\dagger}b_{a,j+1}+h.c.)+U_{12}\sum_{j=1}^{N}n_{1,j}n_{2,j}\nonumber \\
 &  & +\frac{U}{2}\sum_{a=1,2}\sum_{j=1}^{N}n_{a,j}(n_{a,j}-1).\label{eq:hamiltonian}\end{eqnarray}
 We denote the different atomic species with the index $a=1,2$, and
the lattice site with index $j$. We assume that the two species have
the same average filling factor, $\nu=M/N\leq1$, the same intra-species
interaction $U>0$ and hopping parameter $t>0$. The inter-species
interaction is given by $U_{12}$. The operators $b_{a,j}^{\dagger}$
and $b_{a,j}$ are the creation and annihilation operators for atoms
of type $a$ and site $i$ and $n_{a,j}=b_{a,j}^{\dagger}b_{a,j}$
are the number operators.

In Ref. \cite{anzi}, we found that there are four different regimes
in the phase diagram (besides the collapsed or phase-separated regime
that occurs at large $|U_{12}|$): the superfluid (SF), the paired
superfluid (PSF), the counterflow superfluid (CFSF) and the Mott insulator
(MI) state. In addition, each of the superfluid orders can coexist
with a charge density wave (CDW) order. The existence of any such
order is determined from the asymptotic behavior of the correlation
functions of the corresponding order parameters. Specifically, the
single-species superfluid (SF) has the order parameter $b_{a}(x)$
($a=1,2$) and its corresponding correlation function $G(x)=\langle b_{a}^{\dagger}(x)b_{a}(0)\rangle$;
the paired superfluid (PSF) has the order parameter $b_{1}(x)b_{2}(x)$
and its corresponding correlation function $R_{S}(x)=\langle b_{1}^{\dagger}(x)b_{2}^{\dagger}(x)b_{1}(0)b_{2}(0)\rangle$;
the counter-flow superfluid (CFSF) has the order parameter $b_{1}(x)b_{2}^{\dagger}(x)$
and its corresponding correlation function $R_{A}(x)=\langle b_{1}^{\dagger}(x)b_{2}(x)b_{1}(0)b_{2}^{\dagger}(0)\rangle.$
The CDW order parameter is $n_{a}(x)$ ($a=1,2$) and the corresponding
correlation function $R_{n,a}(x)=\langle n_{a}(x)n_{a}(0)\rangle$.
The asymptotic behavior of the correlation functions at large $x$
is listed in Table. \ref{tab:table} for the different phases.

\begin{table}
\begin{tabular}{|c|c|c|c|c|}
\hline 
 & $R_{S}(x)$  & $R_{A}(x)$  & $G(x)$  & $R_{n,a}(x)$\tabularnewline
\hline
\hline 
MI  & E  & E  & E  & A\tabularnewline
\hline 
SF  & A  & A  & A  & A\tabularnewline
\hline 
CFSF  & E  & A  & E  & A\tabularnewline
\hline 
PSF  & A  & E  & E  & A\tabularnewline
\hline 
CDW  & A or E  & A or E  & A or E  & A ($\alpha<2$)\tabularnewline
\hline
\end{tabular}

\caption{\label{tab:table}Definitions of MI, SF, CFSF and PSF orders in terms
of the large $x$ behavior of the correlation functions $R_{S}(x)$,
$R_{A}(x)$, and $G(x)$, $R_{n,a}(x)$. A: algebraic decay of the
form $x^{-\alpha}$; E: exponential decay of the form $e^{-\beta x}$.
A correlation function is said to exhibit quasi-order when it is subject
to algebraic decay with $\alpha<2$. In this system, the algebraic
decay for $R_{S}$, $R_{A}$ and $G$ always has $\alpha<2$, while
$R_{n,a}$ can have $\alpha\geq2$. CDW quasi-order exists only when
$R_{n,a}$ is described by $\alpha<2$. }

\end{table}

We calculate the noise correlations from the four-point correlation
function: \begin{eqnarray}
\mathcal{G}_{aa'}(k,k') & = & \sum_{j_{1},,j_{2},j_{3},j_{4}=1}^{N}\mathcal{L}_{aa'}(j_{1},j_{2},j_{3},j_{4})e^{i[kj_{12}+k'j_{34}]}\nonumber \\
 &  & -\langle n_{a}(k)\rangle\langle n_{a'}(k')\rangle,\label{eq:Gaa}\end{eqnarray}
 where $j_{12}\equiv j_{1}-j_{2}$, $j_{34}\equiv j_{3}-j_{4}$ and
$\mathcal{L}_{aa'}$ is the four-point correlation function, \begin{equation}
\mathcal{L}_{aa'}(j_{1},j_{2},j_{3},j_{4})=\langle b_{a,j_{1}}^{\dagger}b_{a,j_{2}}b_{a',j_{3}}^{\dagger}b_{a'j_{4}}\rangle.\label{eq:four}\end{equation}
 It is easy to see that the correlation functions $R_{S}$, $R_{A}$
and $R_{n,a}$ are the special cases of $\mathcal{L}_{aa'}$,

\begin{eqnarray}
\mathcal{L}_{12}(j_{1},j_{2},j_{1},j_{2}) & = & R_{S}(j_{1},j_{2}),\nonumber \\
\mathcal{L}_{12}(j_{1},j_{2},j_{2},j_{1}) & = & R_{A}(j_{1},j_{2}),\label{eq:LL}\\
\mathcal{L}_{aa}(j_{1},j_{2},j_{2},j_{1}) & = & R_{n,a}(j_{1},j_{2})+n_{a,j_{1}}.\nonumber \end{eqnarray}

The noise correlation $\mathcal{G}_{12}$, therefore, contains the
Fourier transform of $R_{S}$ and $R_{A}$, \begin{equation}
g_{S}(k,k')=\sum_{j_{1},j_{2}}R_{S}(j_{1},j_{2})e^{i(k+k')(j_{1}-j_{2})}\label{eq:gs}\end{equation}
 and\begin{equation}
g_{A}(k,k')=\sum_{j_{1},j_{2}}R_{A}(j_{1},j_{2})e^{i(k-k')(j_{1}-j_{2})}\label{eq:ga}\end{equation}

and $\mathcal{G}_{aa}$ contains the Fourier transform of $R_{n,a}$,

\begin{equation}
g_{n,a}=\sum_{j_{1},j_{2}}R_{n,a}(j_{1},j_{2})e^{i(k-k')(j_{1}-j_{2})}.\label{eq:gn}\end{equation}

If $R_{S}(j_{1},j_{2})$ decays as $|j_{1}-j_{2}|^{-1/K_{S}}$, we
find that $g_{s}$ scales as $|k+k'|^{-1/K_{S}}$. Similarly, if $R_{A}$
decays as $|j_{1}-j_{2}|^{-1/K_{A}}$, $g_{A}$ scales as $|k-k'|^{-1/K_{A}}$.
For the PSF phase, we find that $g_{s}(k,k')$ is the dominant term
of $\mathcal{G}_{12}(k,k')$ with a strong peak around $k=-k'$. This
peak is the signal of the PSF order. Similarly, for the CFSF phase,
we find that the function $g_{A}(k,k')$ becomes dominant around $k=k'$
in $\mathcal{G}_{12}(k,k')$. The peak around $k=k'$ is the signal
of the CFSF order. These remarks are made to give the reader an intuitive
interpretation of the relationship between the noise correlations
and the long-range orders. In the following section, we will explain
the calculation of the noise correlations via LL theory and show that
the features mentioned above are indeed reflected in the LL calculation
results.

\begin{figure*}
\includegraphics[width=4.5cm]{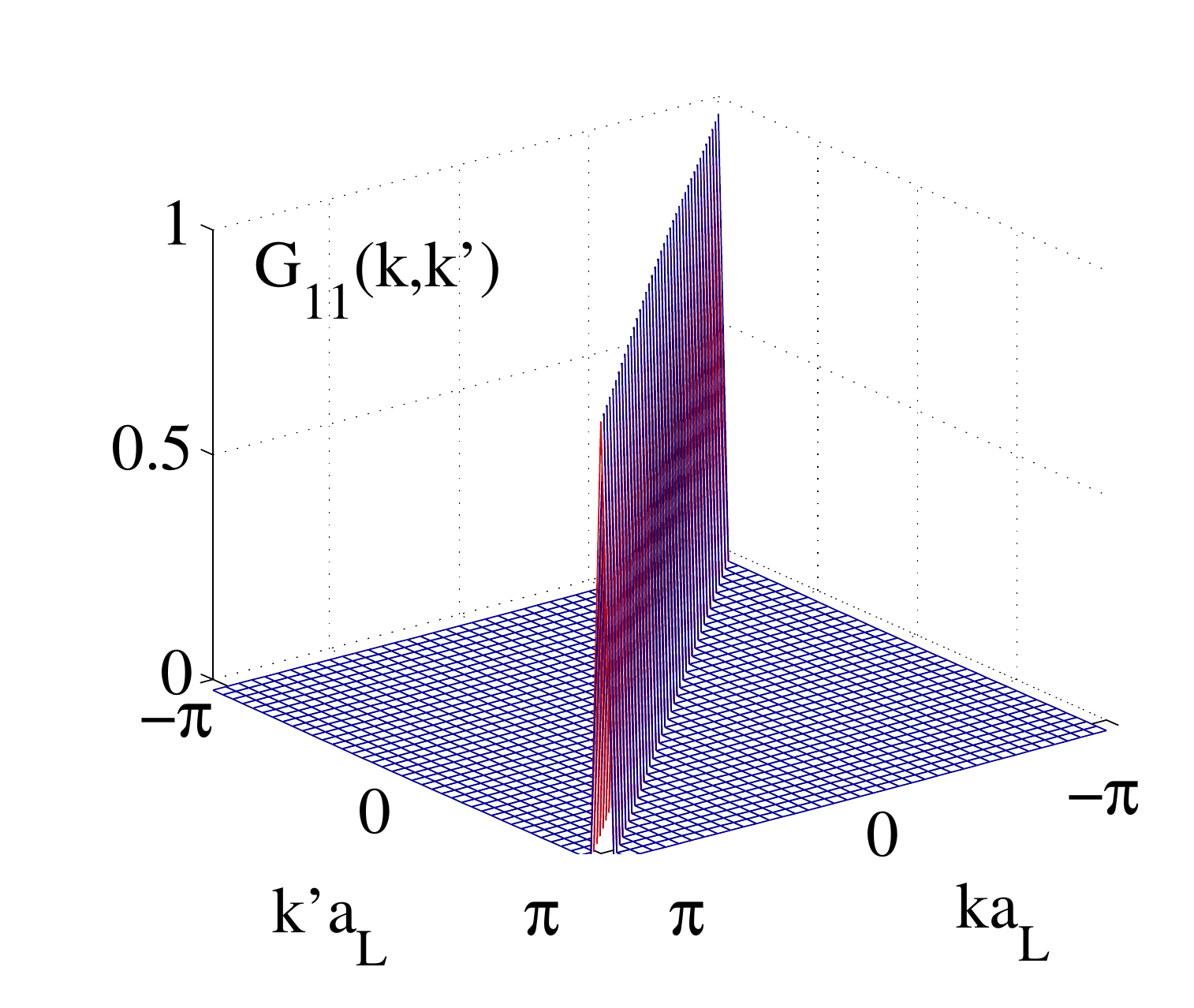}\includegraphics[width=4.5cm]{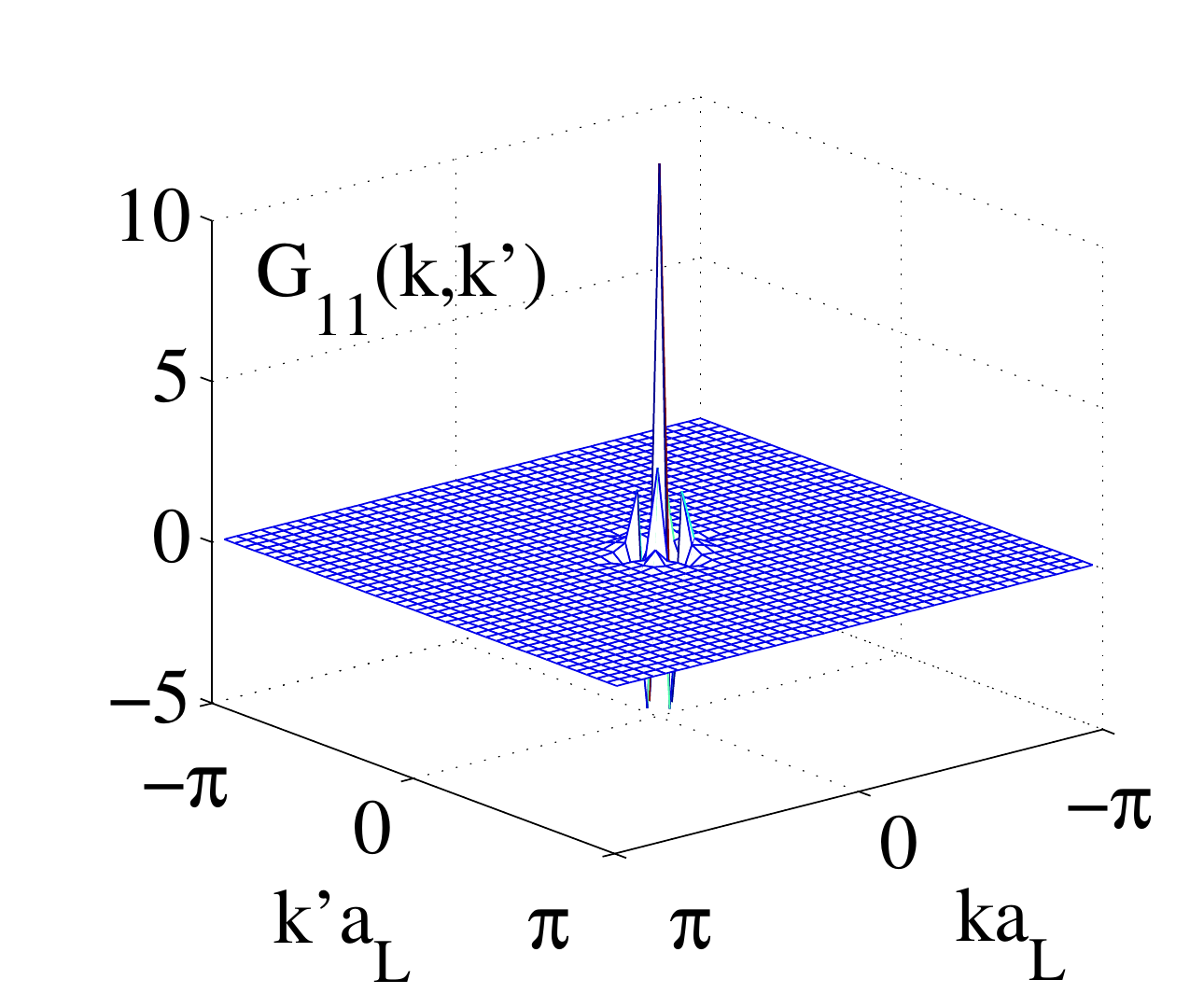}\includegraphics[width=4.5cm]{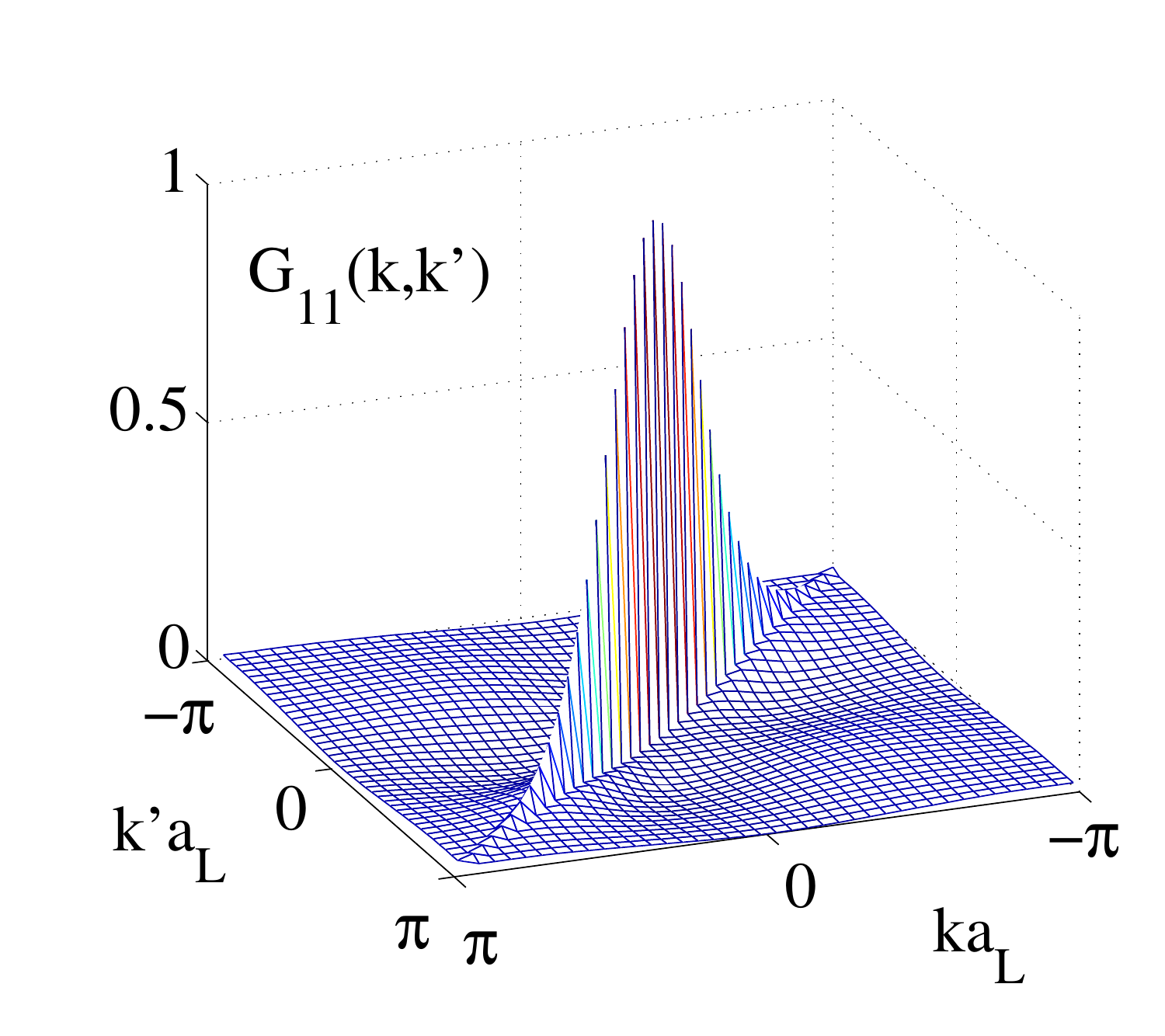}\includegraphics[width=4.5cm]{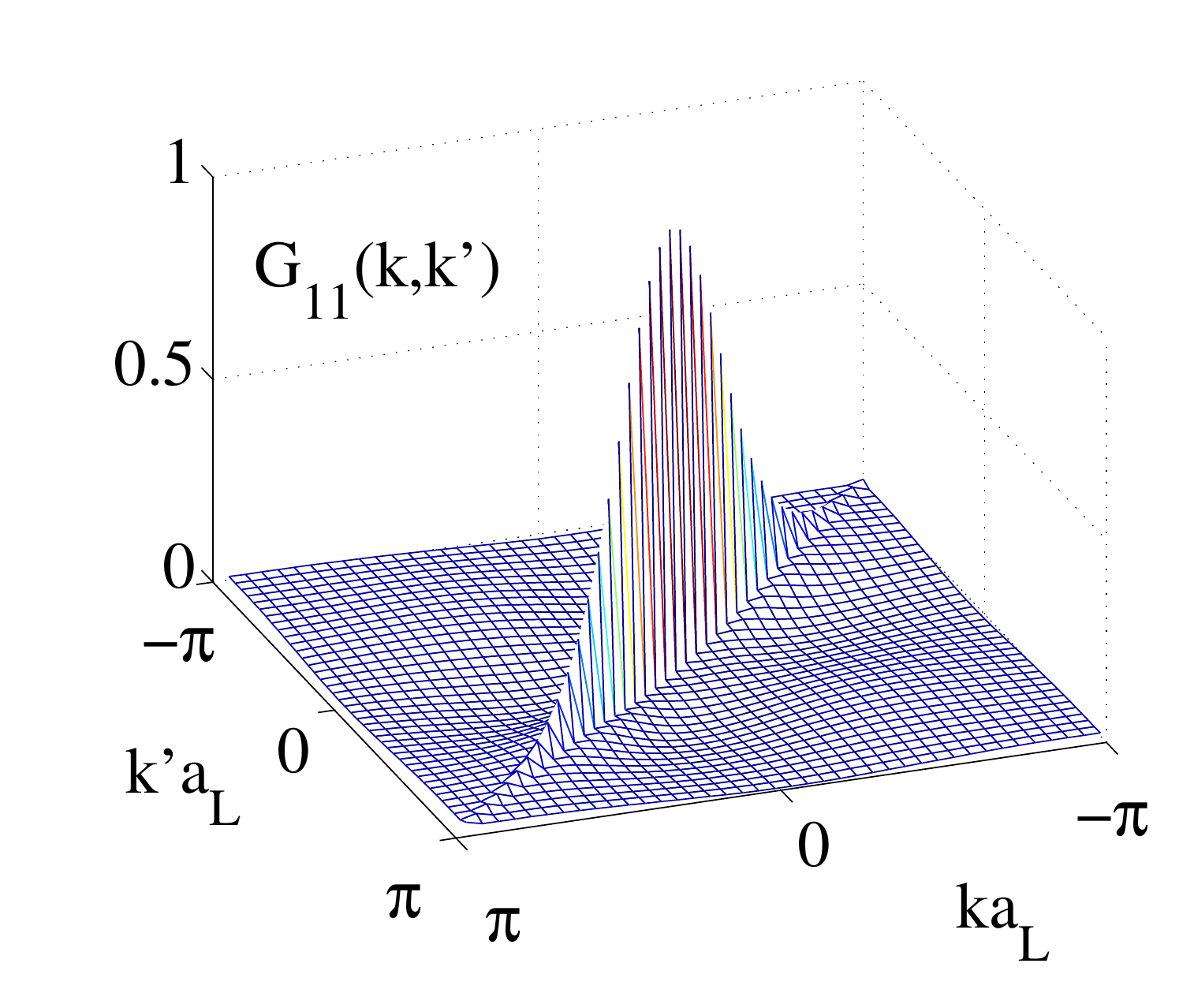}

\subfigure[MI]{\includegraphics[width=4.5cm]{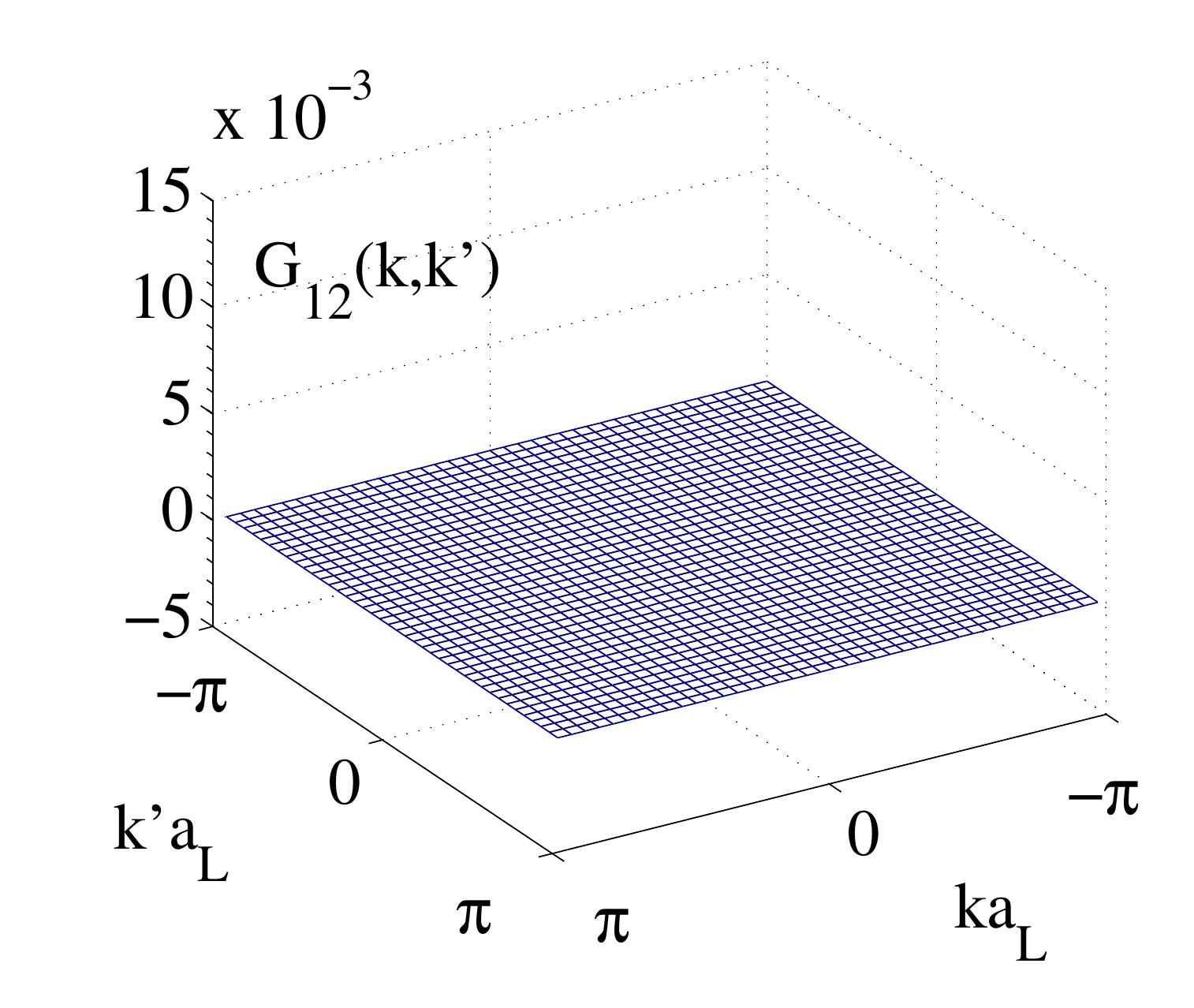}}\subfigure[SF]{\includegraphics[width=4.5cm]{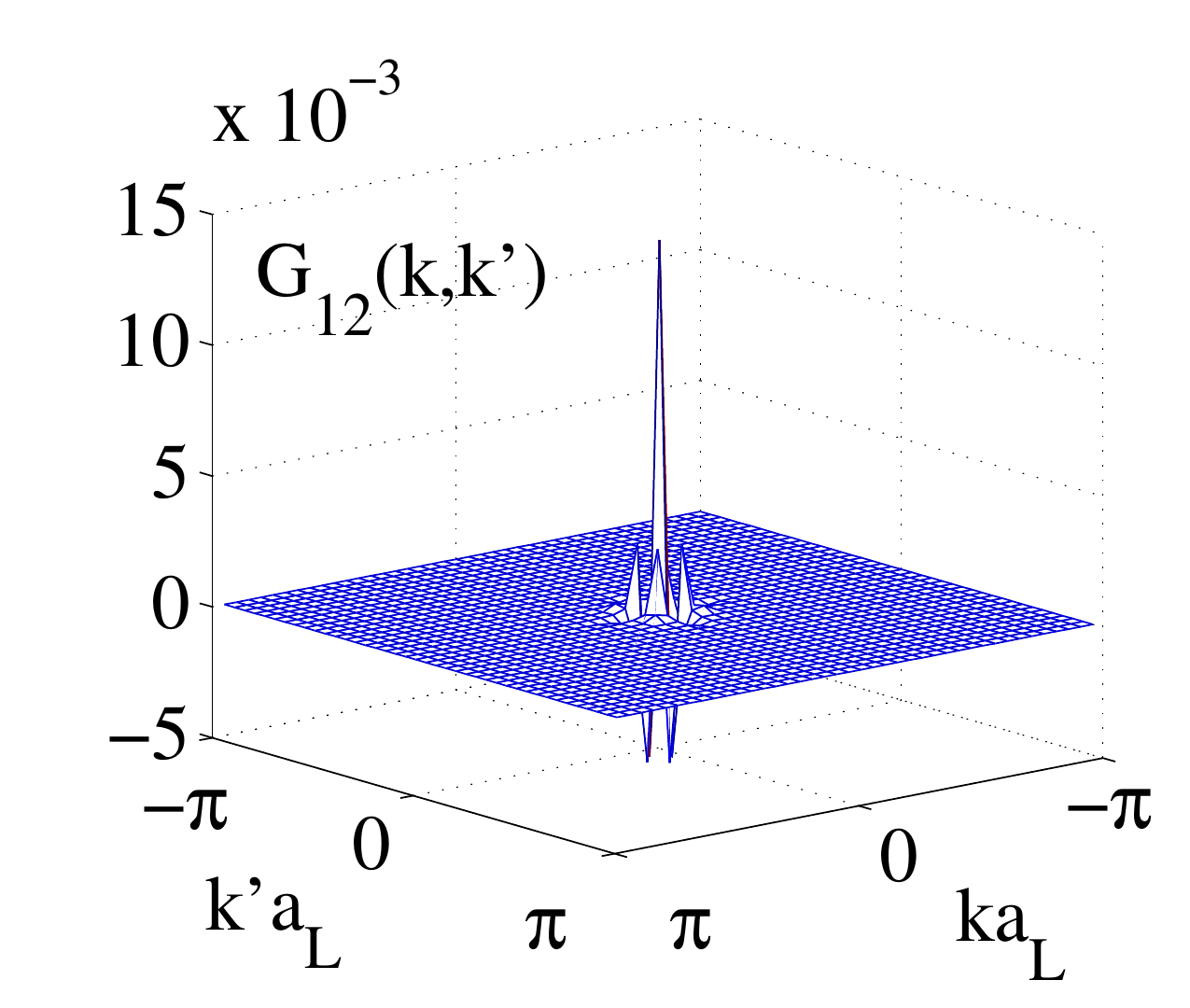}}\subfigure[PSF]{\includegraphics[width=4.5cm]{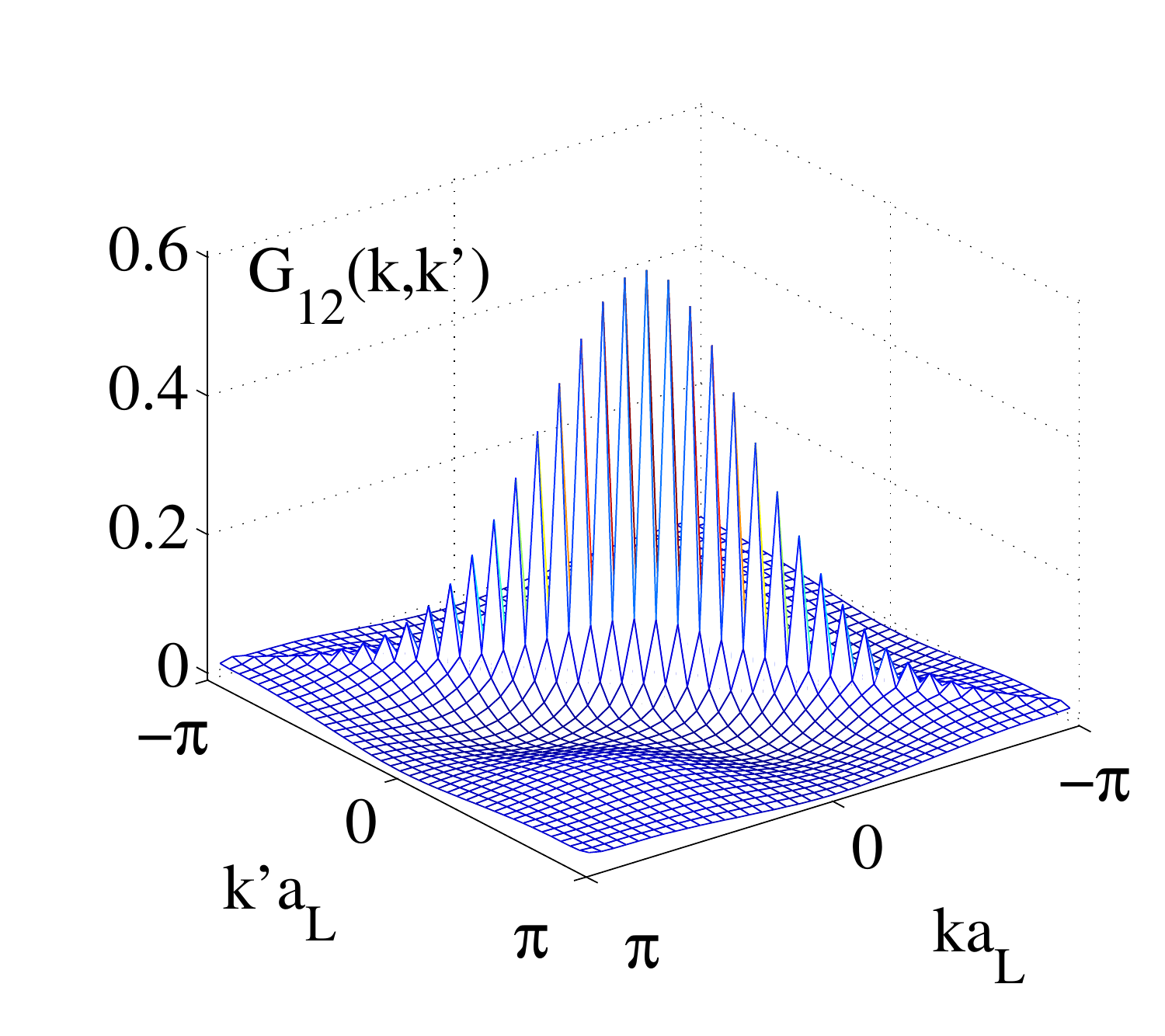}}\subfigure[CFSF]{\includegraphics[width=4.5cm]{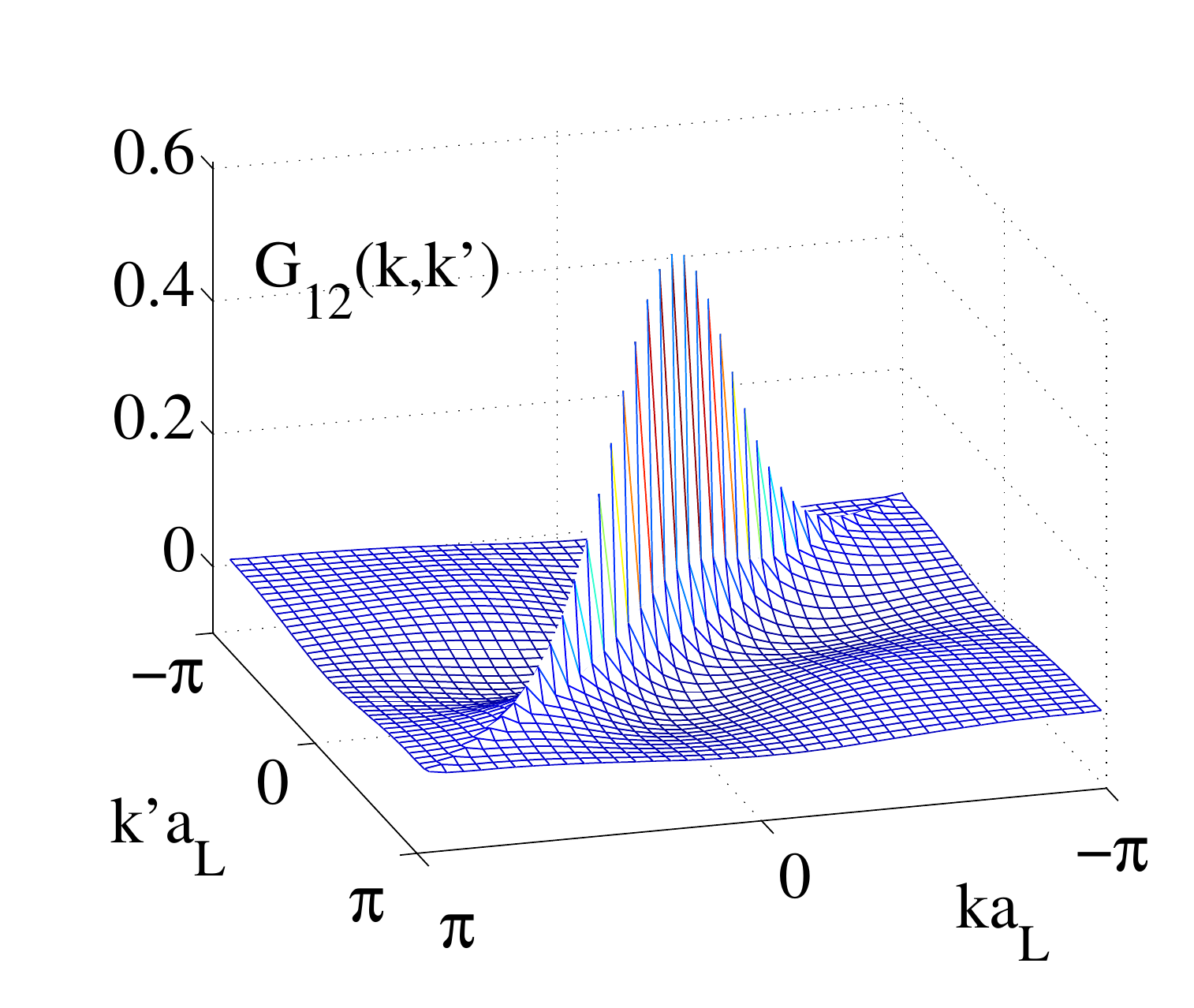}}

\caption{\label{fig:LL} Noise correlations in different phases, derived from
Luttinger liquid theory. In the MI state (column (a)), $\delta$-function
like correlations along $k=k'$ in $\mathcal{G}_{11}(k,k')$ are visible,
whereas $\mathcal{G}_{12}$ nearly vanishes. In the SF state (column
(b)), with Luttinger parameters $K_{A}=1.03$ and $K_{S}=0.96$, we
find various contributions in $\mathcal{G}_{11}$, especially $\delta$-
function along $k=k'$. In Fig. \ref{fig:SF_LLG}, we show the contour
plots for $\mathcal{G}_{11}$ and $\mathcal{G}_{12}$ for the same
state, where we can see the negative correlations at $k=0$ and $k'=0$,
as well as pairing correlation along $k=-k'$, which is similar to
the single-species result in Ref. \cite{LM}. $\mathcal{G}_{12}$
shows similar features, but the bunching contribution is an algebraic
peak, rather than a $\delta$-function. In (c) we show an example
for the PSF phase, with $K_{A}=0.01$ and $K_{S}\simeq1.3$, in (d)
an example for the CFSF phase, with $K_{S}=0.01$ and $K_{A}\simeq1.2$.
In the PSF state, the inter-species correlation $\mathcal{G}_{12}(k,k')$
has strong correlations along $k=-k'$, a reflection of pairing. In
the CFSF state, the peak is formed along $k=k'$ direction, an indication
of the anti-pairing (particle-hole) formation in the CFSF state.}

\end{figure*}

\begin{figure}
\includegraphics[width=4.5cm]{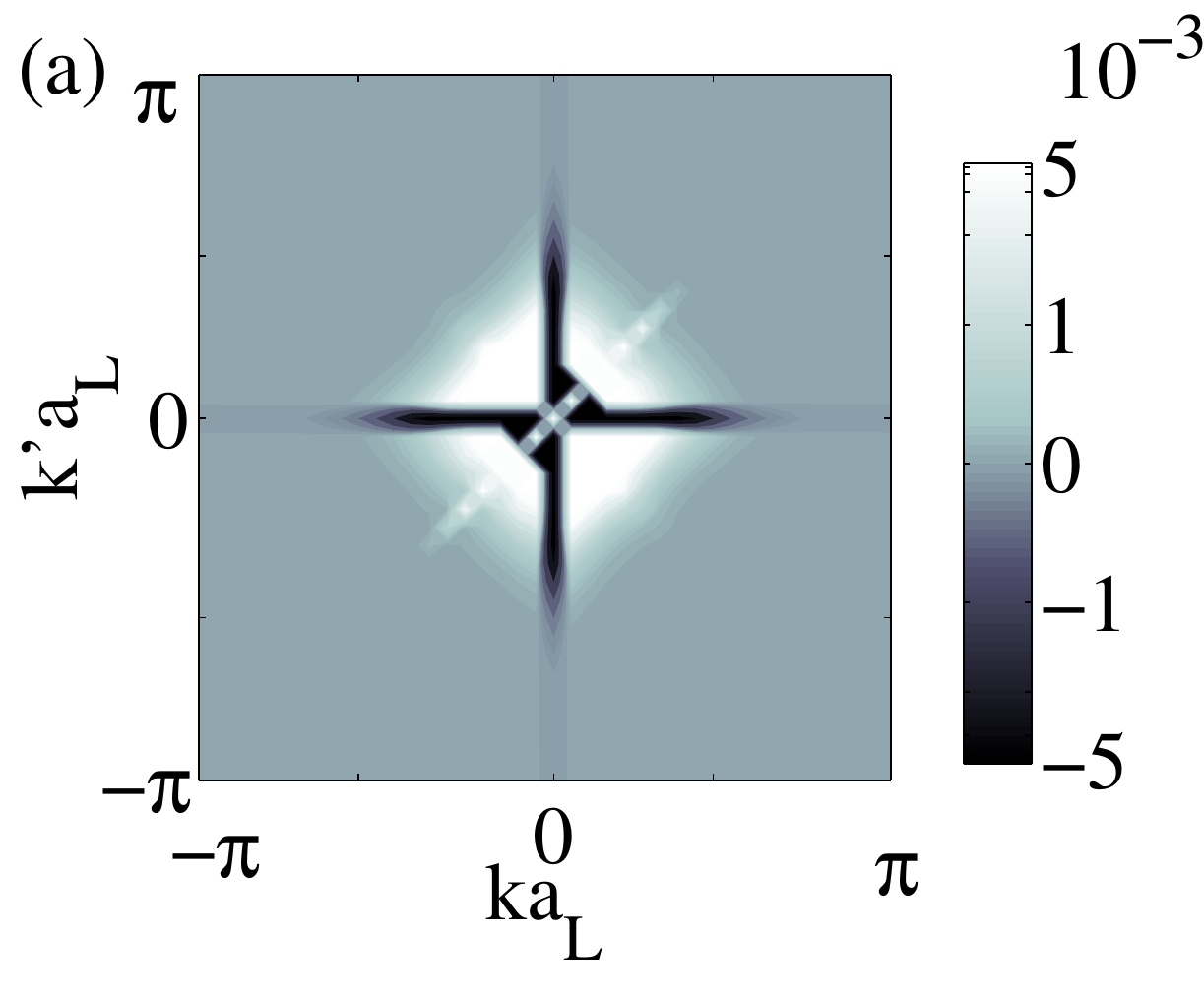}\includegraphics[width=4.5cm]{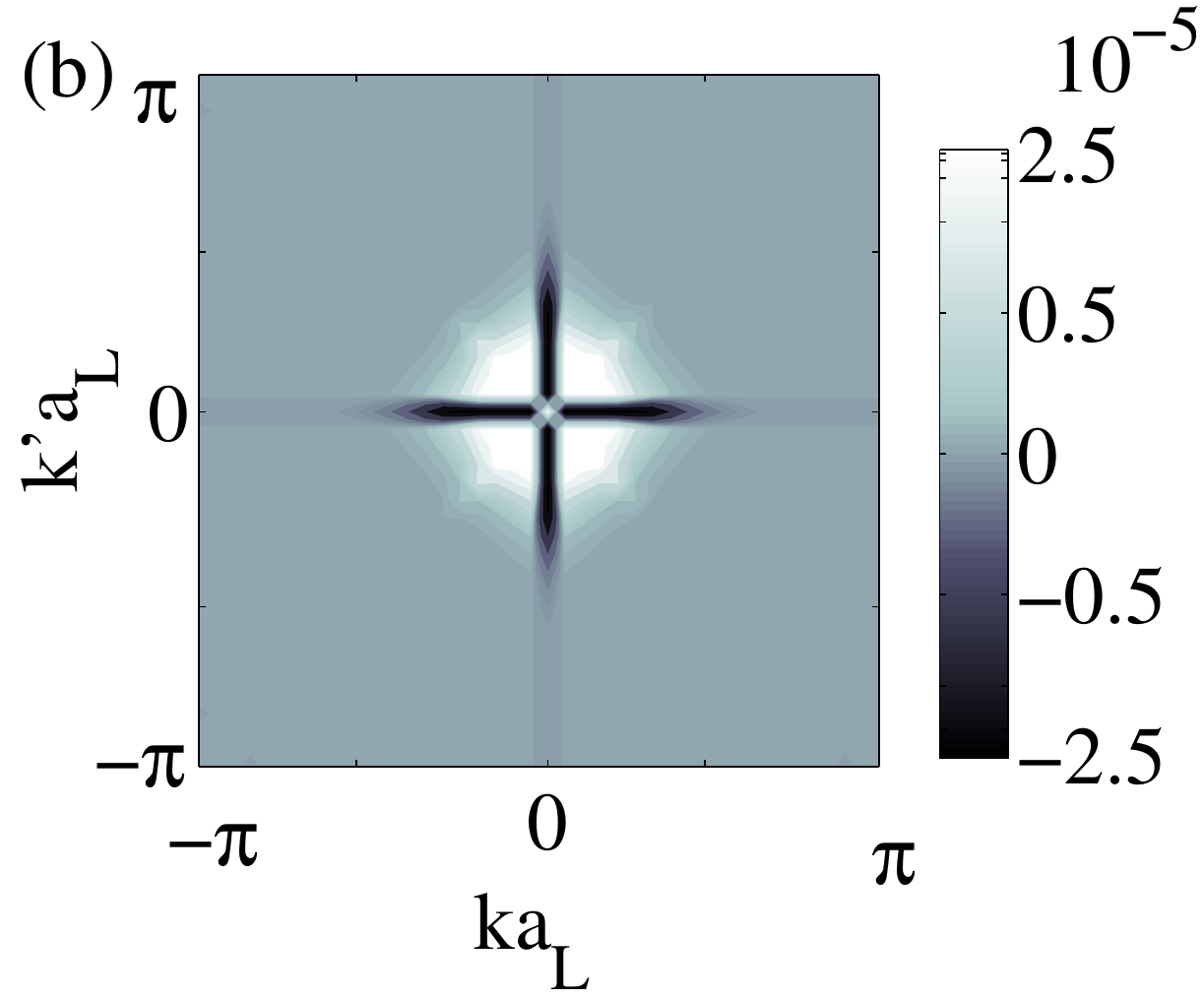}

\caption{\label{fig:SF_LLG}Noise correlations, $\mathcal{G}_{11}(k,k')$ (a)
and $\mathcal{G}_{12}(k,k')$ (b), in the SF state. The data used
here are the same as those used in Fig. \ref{fig:LL} (b). We create
the non-linear gray scales by plotting functions $\tanh(500\mathcal{G}_{11})$
(a) and $\tanh(10^{5}\mathcal{G}_{12})$ (b), in order to magnify
the details around $k=k'=0$. The labels of the color-bar reflect
the values of $\mathcal{G}_{11}(k,k')$ and $\mathcal{G}_{12}(k,k')$.
In these plots, we can clearly see the negative correlations between
the quasi-condensate ($k=0$ and $k'=0$) and the higher momentum
states at the quantum depletion, as well as the anti-pairing correlation
along $k=k'$ and the pairing correlation along $k=-k'$.}

\end{figure}

\section{\label{sec:Luttinger-Liquid-Calculation} Luttinger Liquid Approach}

In this section we determine the generic behavior of the noise correlations
using a Luttinger liquid approach. This formalism has been applied
to one-dimensional Fermi systems in Ref. \cite{LM_Fermion}, and
additionally to single-species bosonic systems in Ref. \cite{LM},
where a detailed description of these calculations was given. Here,
we use an analogous derivation for the case of a bosonic mixture.
We outline key steps of the derivation, but refer the reader to Ref.
\cite{LM} for a detailed description of the method.

As described in Ref. \cite{anzi}, we switch to a continuum description,
in which the single particle operators are given by $b_{a}(x)$. We
then use a bosonization identity \cite{Haldane,Cazalilla} \begin{eqnarray}
b_{a}(x) & = & [n+\Pi_{a}(x)]^{1/2}\sum_{m}e^{2im\Theta_{a}(x)}e^{i\phi_{a}(x)},\end{eqnarray}
 where $n$ is the real space density, related to the filling factor
by $n=\nu/a_{L}$, where $a_{L}$is the lattice constant, and $m$
is an integer summation index. For future reference, we note that
the Fermi wavevector of an equivalent system of fermions, $k_{F}$,
is given by $k_{F}=\pi n$. Although this paper describes a bosonic
system, we find that the Fermi wavevector occurs naturally in a number
of contexts. For example, $\Theta_{a}$ is given by $\Theta_{a}(x)=k_{F}x+\theta_{a}(x)$,
where $\theta_{a}(x)=\pi\int^{x}dy\Pi_{a}(y)$. $\Pi_{a}$ describes
the low-momentum density fluctuations of species $a$ and $\phi_{a}(x)$
is the phase field of species $a$.

We calculate the noise correlations at the Gaussian fixed point, corresponding
to the SF phase. Here, the system separates into symmetric and anti-symmetric
degrees of freedom, defined as $\theta_{S,A}=(\theta_{1}\pm\theta_{2})/\sqrt{2}$
and $\phi_{S,A}=(\phi_{1}\pm\phi_{2})/\sqrt{2}$. The action can be
written either in terms of the phase fields \begin{eqnarray}
S & = & \sum_{j=S,A}\int d^{2}r_{j}\Big[\frac{K_{j}}{2\pi}[(\partial_{v_{j}\tau}\phi_{j})^{2}+(\partial_{x}\phi_{j})^{2}]\Big],\end{eqnarray}
 or in terms of the fields $\theta_{S,A}$ \begin{eqnarray}
S & = & \sum_{j=S,A}\int d^{2}r_{j}\Big[\frac{1}{2\pi K_{j}}[(\partial_{v_{j}\tau}\theta_{j})^{2}+(\partial_{x}\theta_{j})^{2}]\Big].\end{eqnarray}
 The velocities $v_{S,A}$ are the phonon velocities of the symmetric/anti-symmetric
modes, and ${\bf r}_{S,A}=(v_{S,A}\tau,x)$. The parameters $K_{S,A}$
are the Luttinger parameters of the symmetric/anti-symmetric sector.
To calculate the noise correlations away from the SF regime, we take
the limits $K_{S,A}\rightarrow0$ to describe the phases in which
either or both $R_{S/A}$ have short-ranged correlations (exponential
decay). This approximation corresponds to the limit that the length
scale of the exponential decay is much smaller than any other length
scale of the system. Our calculation could be extended in a straightforward
way to include a finite decay length of the exponential decay.

We start out by calculating $\langle n_{a,k}\rangle$ for small momentum
$k\approx0$, for which the Bose operators are given by $b_{a}\sim\sqrt{n}e^{i\phi_{a}}$.
For $\langle n_{a,k}\rangle$ we find: \begin{eqnarray}
\langle n_{a,k}\rangle & \sim & n\int dx_{12}e^{ikx_{12}}e^{-\frac{1}{2}\langle(\phi_{a}(2)-\phi_{a}(1))^{2}\rangle},\end{eqnarray}
 where $\phi_{a}(1)$ refers to $\phi_{a}(x_{1})$, and similarly
for $\phi_{a}(2)$, and $x_{12}=x_{1}-x_{2}$. The correlation function
$\langle(\phi_{a}(2)-\phi_{a}(1))^{2}\rangle$ can be rewritten in
terms of correlation functions for $\phi_{S,A}$. Using the Gaussian
action above, we find \begin{eqnarray}
\langle(\phi_{S/A}(2)-\phi_{S/A}(1))^{2}\rangle & = & \frac{1}{2K_{S/A}}\log\frac{r_{0}^{2}+x_{12}^{2}}{r_{0}^{2}},\end{eqnarray}
 where $r_{0}$ is a short-range cut-off. With that we find \begin{eqnarray}
\langle n_{k}\rangle\sim n\int dx_{12}e^{ikx_{12}}\mathcal{F}(x_{12}),\label{nk}\end{eqnarray}
 where \begin{eqnarray}
\mathcal{F}(x) & =\left(\frac{r_{0}^{2}}{r_{0}^{2}+x^{2}}\right)^{g}.\end{eqnarray}
 The exponent $g$ is given by $g=1/8K_{S}+1/8K_{A}$. Next we evaluate
the expectation value $\langle n_{k}n_{k'}\rangle$ along the same
lines. We obtain: \begin{eqnarray}
\langle n_{1,k}n_{1,k'} & \rangle\sim & n^{2}\int e^{ikx_{12}+ik'x_{34}}\mathcal{F}(x_{12})\mathcal{F}(x_{34})\mathcal{A},\label{nnBos}\end{eqnarray}
 where \begin{eqnarray}
\mathcal{A} & =\left(\frac{(r_{0}^{2}+x_{14}^{2})(r_{0}^{2}+x_{23}^{2})}{(r_{0}^{2}+x_{13}^{2})(r_{0}^{2}+x_{24}^{2})}\right)^{h},\label{Ah}\end{eqnarray}
 and Eq. \ref{nnBos} is a volume integral over the spatial variables
$x_{12}$, $x_{23}$, $x_{34}$. The exponent $h$ is given by $h=-1/8K_{S}-1/8K_{A}$.
We combine these expressions to get the correlation function $\mathcal{G}_{11}(k,k')$:

\begin{eqnarray}
 &  & \mathcal{G}_{11}(k,k')\nonumber \\
 &  & \sim n^{2}\int e^{ikx_{12}+ik'x_{34}}\mathcal{F}(x_{12})\mathcal{F}(x_{34})(\mathcal{A}-1).\label{eq:noise_B}\end{eqnarray}
 For $\mathcal{G}_{12}(k,k')$ we proceed analogously, and find $h=-1/8K_{S}+1/8K_{A}$.
For the finite-size systems that we treat here, we evaluate these
integrals numerically, by choosing a finite length $L$ of the system,
and by replacing each spatial variable $x$ by $(L/2\pi)\sin(2\pi x/L)$
(see Ref. \cite{giamarchi_book}). To compare with the TEBD calculations
of a homogeneous system in the next section, we choose the values
of the Luttinger parameters, $K_{A}$ and $K_{S}$ to those obtained
from the TEBD calculations and they are listed in Table. \ref{tab:param}.

In Fig. \ref{fig:LL} (b) and \ref{fig:SF_LLG}, we show an example
for the SF regime. In the upper panel of Fig. \ref{fig:LL} (b), we
show $\mathcal{G}_{11}(k,k')$, in the lower panel $\mathcal{G}_{12}(k,k')$.
The Luttinger parameters are $K_{A}=1.03$ and $K_{S}=0.96$. The
ratio $L/r_{0}$ was chosen as $L/r_{0}=20$, corresponding to the
particle number of each species in the numerical example. The shape
of $\mathcal{G}_{11}(k,k')$ is the same as the noise correlation
function for a single bosonic SF, which was discussed in Ref. \cite{LM}.
It has the characteristic features of a superfluid: positive correlations
along $k=-k'$, which indicates pairing correlations; negative correlations
for the axes $k=0$ and $k'=0$, indicating the negative correlations
between the quasi-condensate and the higher momenta due to pair fluctuations;
and bunching correlations along $k=k'$. For $\mathcal{G}_{12}(k,k')$
we find qualitatively a similar shape, with the main difference, that
the bunching along $k=k'$ does not have a $\delta$-function contribution,
but only algebraic terms. We note that for a system of two non-interacting
species, i.e. for $U_{12}=0$, $K_{S}=K_{A}$, and $\mathcal{G}_{12}$
vanishes.

In Figs. \ref{fig:LL} (c) and (d) we show the noise correlations
for the PSF and the CFSF phase, respectively. For the PSF example,
the Luttinger parameters are $K_{S}=1.3$ and $K_{A}=0.01$. For $L/r_{0}$
we again pick $L/r_{0}=20$. For the CFSF phase, the parameters are
$K_{A}=1.2$ and $K_{S}=0.01$. In the PSF regime, we find a strong
pairing signature in $\mathcal{G}_{12}$, similar to the pairing signature
in Fermi mixtures \cite{LM_Fermion,LM}. In the CFSF example, an
strong anti-pairing signature is found in $\mathcal{G}_{12}$.

We can obtain the functional form of these signatures in the limit
$L\rightarrow\infty$, by applying similar arguments to what has been
given in Ref. \cite{LM}. In the PSF region, We rewrite the noise
correlation integral in terms of $z=(x_{12}-x_{34})/2$, $h_{+}=(x_{14}+x_{23})/2$
and $h_{-}=(x_{14}-x_{23})/2$. We then note that for $\mathcal{G}_{12}$
and for $K_{A}\rightarrow0$, the exponent $h=-1/8K_{S}+1/8K_{A}$
diverges. This enforces the integrand to be negligible away from $z$,
$h_{+}\approx0$. Thus the integral evaluates to \begin{eqnarray}
\mathcal{G}_{12} & \sim & |k+k'|^{-1/K_{S}}.\label{eq:G_PSF}\end{eqnarray}
 This is the shape that would be approached in an infinite system
by the noise correlations shown in Fig. \ref{fig:LL} (c), lower panel.
The deviation from the pure power law is due to the finite size of
the system. With similar arguments one can show that in the CFSF regime
the inter-species noise correlation approaches \begin{eqnarray}
\mathcal{G}_{12} & \sim & |k-k'|^{-1/K_{A}},\label{eq:G_CFSF}\end{eqnarray}
 for $L\rightarrow\infty$. Again, the deviation from a pure power
law is due to the finite size of the system. Furthermore, one can
show that for both PSF and CFSF orders, $\mathcal{G}_{11}(k,k')$
approaches $\delta(k-k')$, in the limit of infinite size. Equations
\ref{eq:G_PSF} and \ref{eq:G_CFSF} show that there is a simple relationship
between $\mathcal{G}_{12}$ and $K_{S}$ in the PSF regime and $\mathcal{G}_{12}$
and $K_{A}$ in the CFSF regime. This suggests that a careful measurement
of $\mathcal{G}_{12}$ can be used to extract the value of the Luttinger
parameters appropriate to the system. This is further confirmed by
our numerical calculations for a trapped system, where we show that
the algebraic relationship described by Eqs. \ref{eq:G_PSF} and \ref{eq:G_CFSF}
remains valid in the presence of a harmonic trap. We discuss prospects
for experimental determination of Luttinger parameters in Sec. \ref{sub:Determination}.

The MI result in Fig. \ref{fig:LL} (a) is obtained by setting both
$K_{A}$ and $K_{S}$ to 0.01. In this case, the ground state closely
approximates a simple product of MI states of each species. Thus,
$\mathcal{G}_{11}$ approaches a $\delta$-function, whereas $\mathcal{G}_{12}$
nearly vanishes.

Next we calculate the noise correlations for the case $k\approx0$
and $k'\approx2k_{F}$, where $k_{F}$ is the Fermi wavevector defined
above. Essentially the same calculation can be done for $k'\approx-2k_{F}$,
and $k'\approx0$ and $k\approx\pm2k_{F}$. $n_{a,k}$ is still given
by the expression (\ref{nk}), but $n_{a,k'}$ now needs to be calculated
with the operator representation $b(x)=\sqrt{n}\exp(2i\Theta(x))\exp(i\phi(x))$.
With that we find \begin{eqnarray}
\langle n_{q'+2k_{F}}\rangle\sim n\int dx_{12}e^{iq'x_{12}}\mathcal{F}'(x_{12}),\label{nkp}\end{eqnarray}
 where $\mathcal{F}'(x_{12})$ has the same form as before but with
an exponent $g'=1/8K_{S}+1/8K_{A}+(K_{S}+K_{A})/2$. The noise correlations
take the form

\begin{eqnarray}
 &  & \mathcal{G}_{11}(k,q')\nonumber \\
 &  & \sim n^{2}\int e^{ikx_{12}+iq'x_{34}}\mathcal{F}(x_{12})\mathcal{F}'(x_{34})(\mathcal{A}-1).\label{eq:noise_B2}\end{eqnarray}
 We therefore note that around the points $k\approx0$ and $k'\approx\pm2k_{F}$,
and $k\approx\pm2k_{F}$ and $k'\approx0$ the integrand is multiplied
by a contribution that is of the form of the integrand of the static
structure factor \begin{eqnarray}
S(q) & \sim & \int e^{iqx_{12}}\left(\frac{r_{0}^{2}}{r_{0}^{2}+x_{12}^{2}}\right)^{(K_{S}+K_{A})/2},\end{eqnarray}
 which can create cusps in the noise correlation when the system is
in the CDW regime. These cusps are found in our numerical calculations
and are discussed in the next section.

\begin{table*}[t]
\begin{tabular}{|l|c|c|}
\hline 
Parameter setting  & Order  & Luttinger Parameters\tabularnewline
\hline
\hline 
(a) $U_{12}/U=0.01$, $\nu=1$  & Mott Insulator (MI)  & $K_{A}=K_{S}=0$\tabularnewline
\hline 
(b) $U_{12}/U=0.01$, $\nu=0.5$  & Superfluid (SF)  & $K_{A}\simeq1.03$, $K_{S}\simeq0.96$\tabularnewline
\hline 
(c) $U_{12}/U=-0.11$, $\nu=0.5$  & Paired Superfluid (PSF)  & $K_{A}=0$, $K_{S}\simeq1.3$\tabularnewline
\hline 
(d) $U_{12}/U=0.11$, $\nu=0.5$  & Counterflow Superfluid with (CFSF)  & $K_{A}\simeq1.2$, $K_{S}=0$\tabularnewline
\hline
\hline 
(e) $U_{12}/U=0.26$, $\nu=0.2$  & Superfluid with charge density wave (SF/CDW)  & $K_{A}\simeq1.4,$ $K_{S}\simeq0.57$\tabularnewline
\hline
\end{tabular}

\caption{\label{tab:param}The parameters used in the numerical examples and
the Luttinger parameters extracted from the algebraic fit of correlation
functions, $R_{A}$ and $R_{S}$. The Luttinger parameters are set
to zero when the correlations decay exponentially. The hopping parameter
$t$ is $0.02U$ for all cases. The parameters are chosen to represent
different orders that can exist in this system. }

\end{table*}

\section{\label{sec:Numerical-Calculation}Numerical Calculation}

The calculation of noise correlations is based on the ground state
generated by the time-evolving block decimation method (TEBD). This
method has been used to generate the ground state of many 1D models
\cite{Vidal}. In this method, the Hilbert space $\mathbf{H}$ is
decomposed as \begin{equation}
\mathbf{H}=\otimes_{l=1}^{N}\mathbf{H}_{l}.\end{equation}
 Here, $l$ refers to the $l$th lattice site, $N$ is the number
of sites, and $\mathbf{H}_{l}$ is the local Hilbert space at site
$l$ with local dimension $d$, independent of $l$. Any state $|\Psi\rangle$
in $\mathbf{H}$ is represented as \begin{eqnarray}
|\Psi\rangle=\sum_{j_{1},j_{2},\ldots,j_{M}=1}^{d}c_{j_{1},j_{2},\ldots,j_{M}}|j_{1}\rangle|j_{2}\rangle\cdots|j_{M}\rangle,\end{eqnarray}
 where \begin{eqnarray}
c_{j_{1},j_{2},\ldots,j_{M}}\!\!\! & = & \!\!\!\sum_{\alpha_{1}=1}^{\chi}\sum_{\alpha_{2}=1}^{\chi}\cdots\sum_{\alpha_{M-1}=1}^{\chi}\lambda_{\alpha_{1}}^{[1]}\Gamma_{\alpha_{1}\alpha_{2}}^{[1]j_{1}}\lambda_{\alpha_{2}}^{[2]}\Gamma_{\alpha_{2}\alpha_{3}}^{[2]j_{2}}\lambda_{\alpha_{3}}^{[3]}\cdots\nonumber \\
 &  & \times\lambda_{\alpha_{M-1}}^{[M-1]}\Gamma_{\alpha_{M-1}\alpha_{M}}^{[M-1]j_{M-1}}\lambda_{\alpha_{M}}^{[M]}\Gamma_{\alpha_{M-1}\alpha_{M}}^{[M]j_{M}}\lambda_{\alpha_{M+1}}^{[M+1]}.\label{eq:tensorproduct}\end{eqnarray}
 The variables $\lambda_{\alpha_{l}}^{[l]}$ and $\chi_{l}$ are the
Schmidt coefficients and rank of the Schmidt decomposition of $|\Psi\rangle$
at site $l$ and $\Gamma^{[l]}$ is a rank-three tensor. Further detail
on this method is provided in the appendix of the previous publication
\cite{anzi}. Here, we limit ourselves to stating values of parameters
and particular methods of calculation. In this work, we set the Schmidt
rank $\chi=100$ and the local dimension $d=5$. We use imaginary-time
propagation to generate the ground state. After obtaining the ground
state, we calculate the correlation functions, $R_{A}$, $R_{S}$,
$R_{n,a}$ and $G$, and determine the quasi-long range order present
in the system based on the relationship shown in Table. \ref{tab:table}.
Furthermore, we can extract the value of the Luttinger parameters,
$K_{A}$ and $K_{S}$, from the numerically calculated correlation
functions \cite{anzi}. We use these parameters in a LL calculation
to compare the numerical and the analytical results.

The main challenge of determining the noise correlation functions
is the high computational cost of calculating the four-point function,
$\mathcal{L}_{aa'}(j_{1},j_{2},j_{3},j_{4})$ (Eq. \ref{eq:four}),
which is estimated to scale as $\chi^{3}d^{3}N^{4}$. For the system
sizes used in this paper, we use parallel computing algorithms to
speed up the calculation by parallelization the computation of $\mathcal{L}_{aa'}$
along the indices $j_{i}$.

\begin{figure*}[t]
\includegraphics[width=4.5cm]{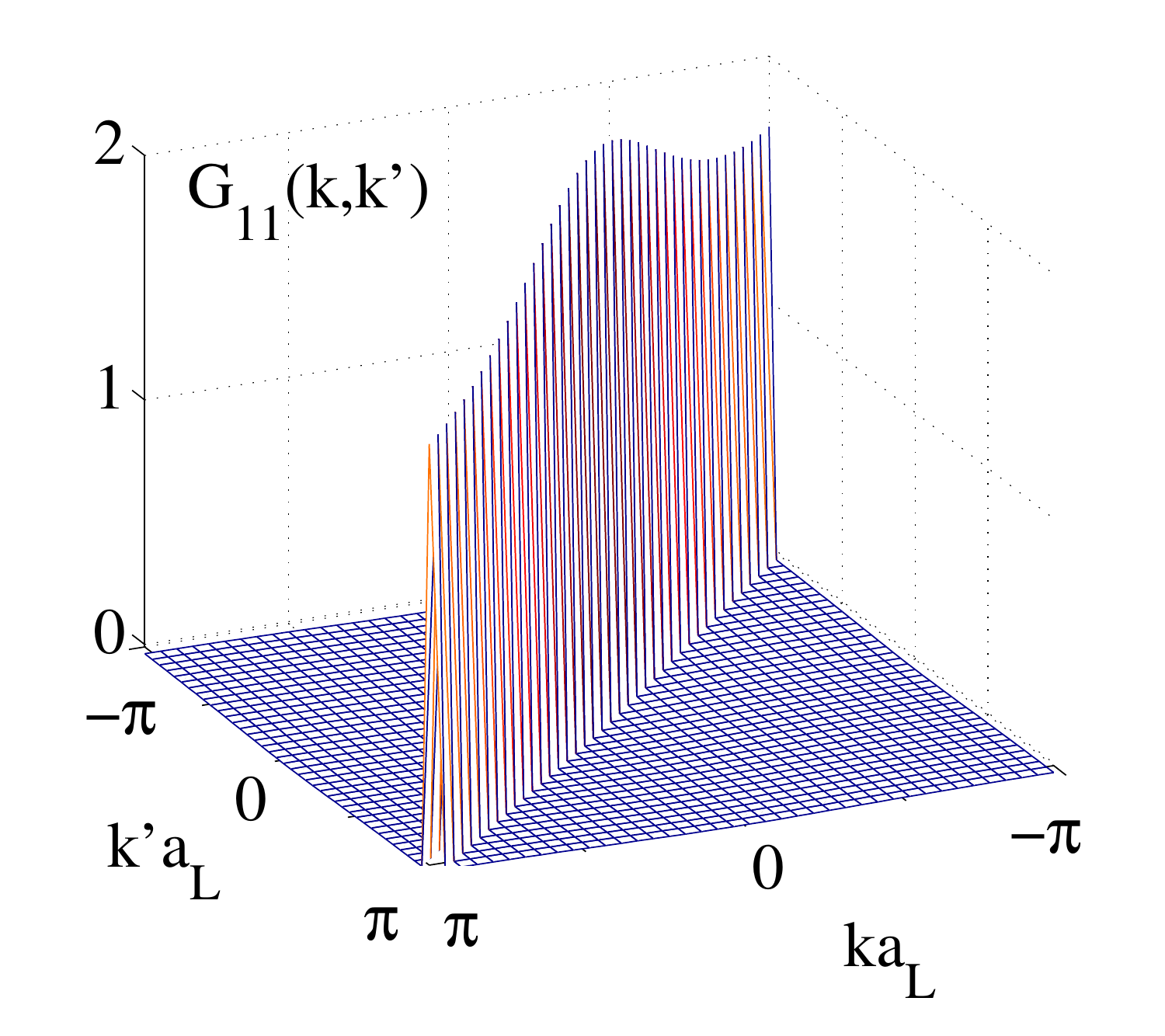}\includegraphics[width=4.5cm]{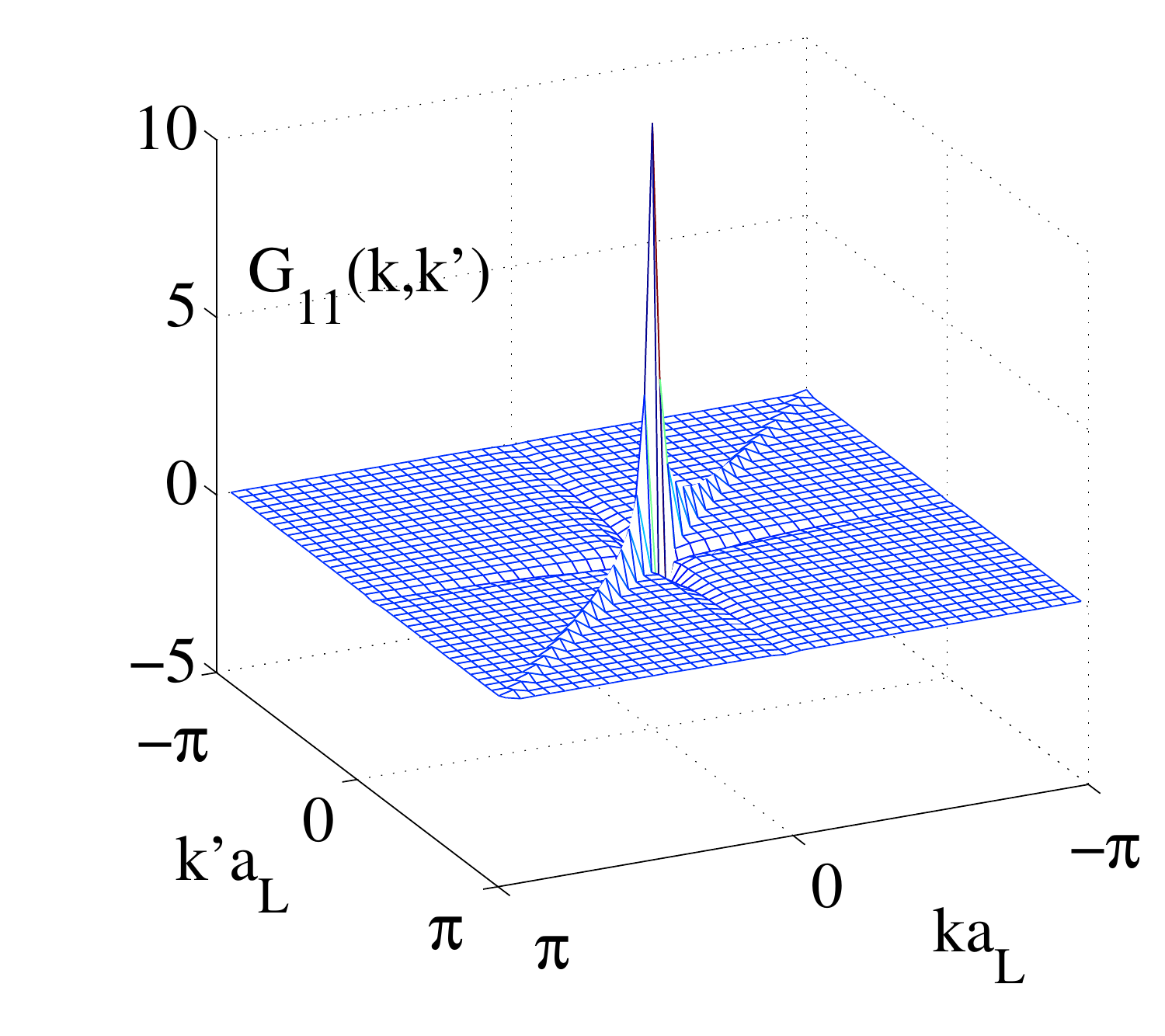}\includegraphics[width=4.5cm]{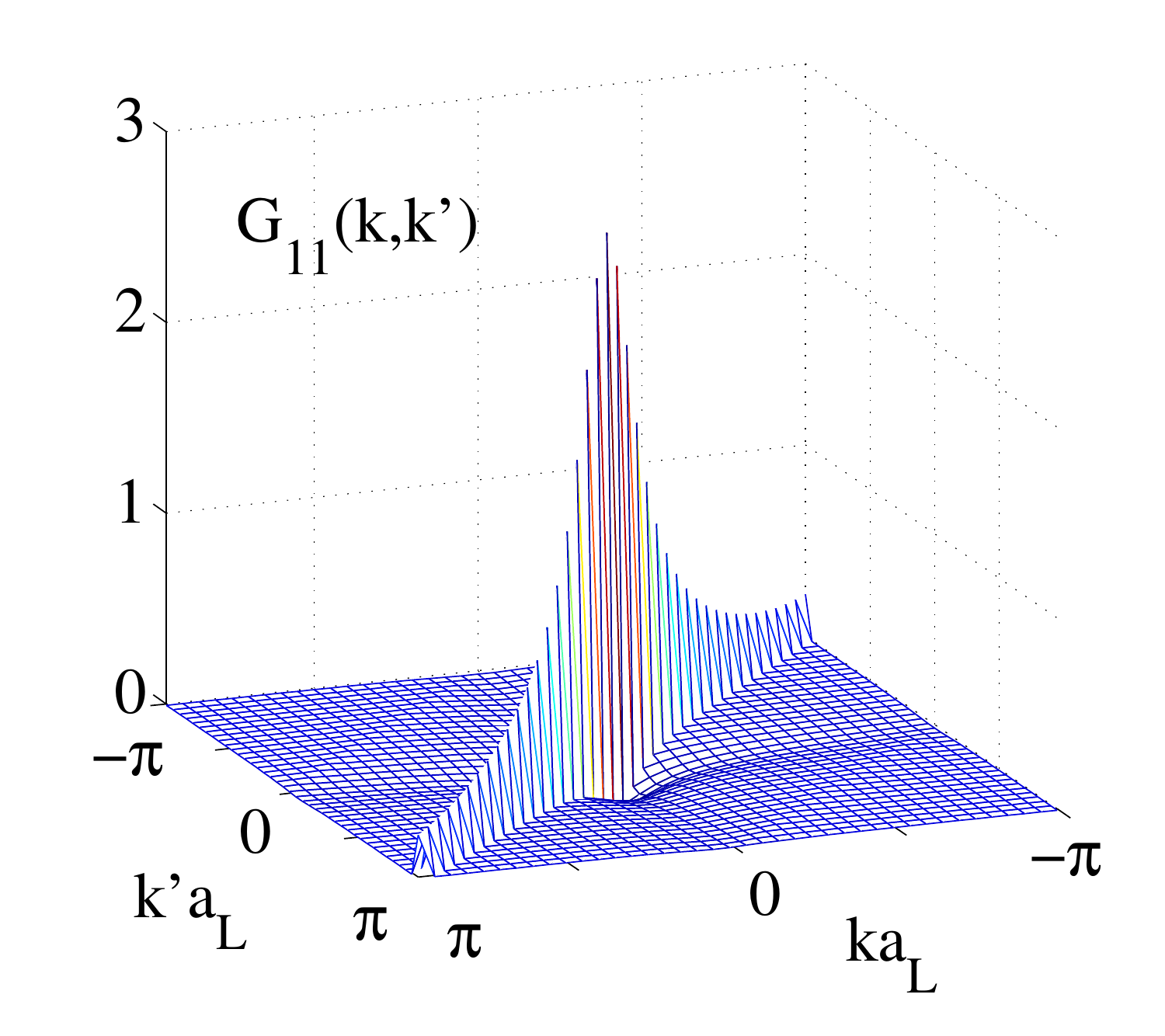}\includegraphics[width=4.5cm]{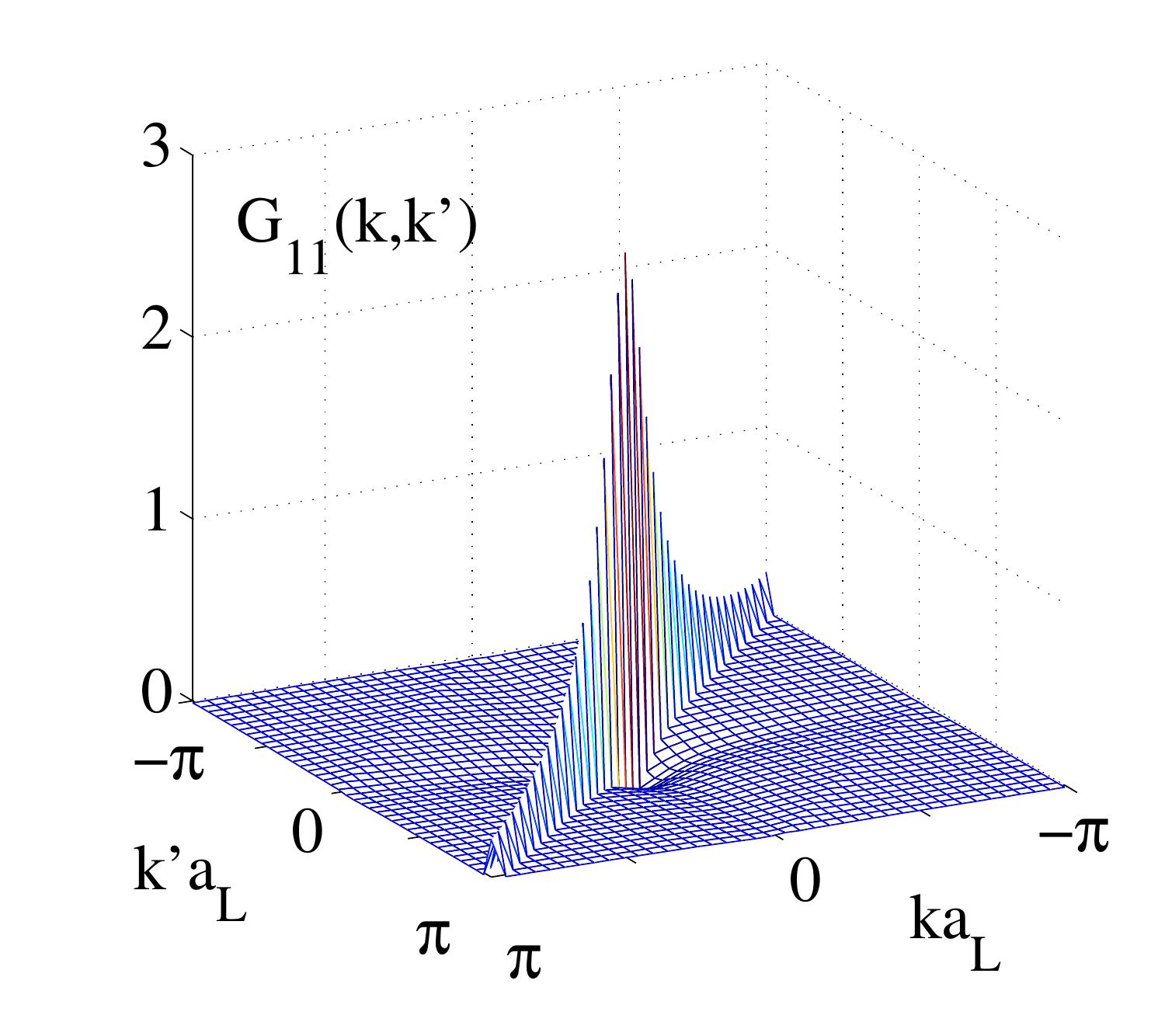}

\subfigure[MI]{\includegraphics[width=4.5cm]{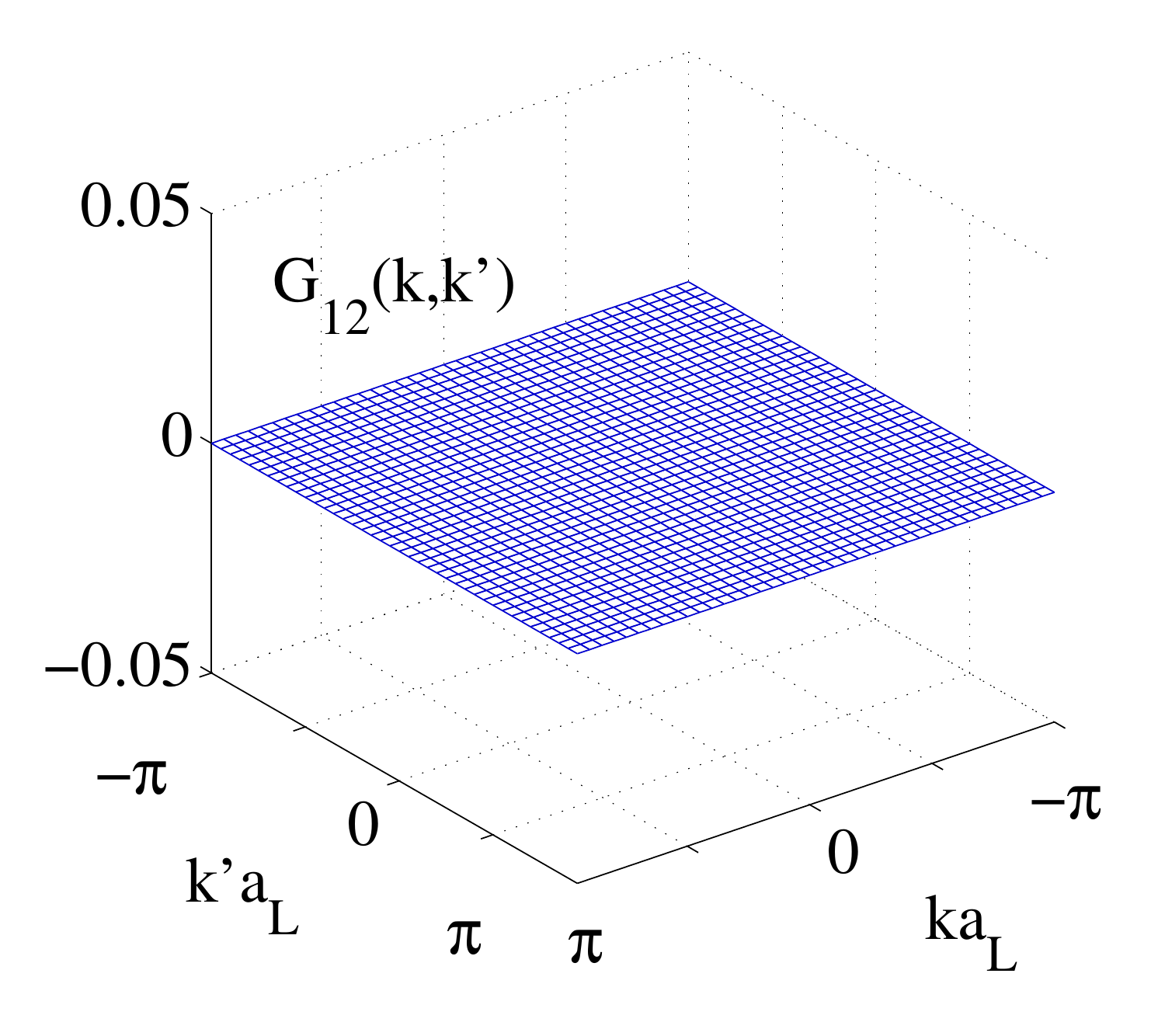}}\subfigure[SF]{\includegraphics[width=4.5cm]{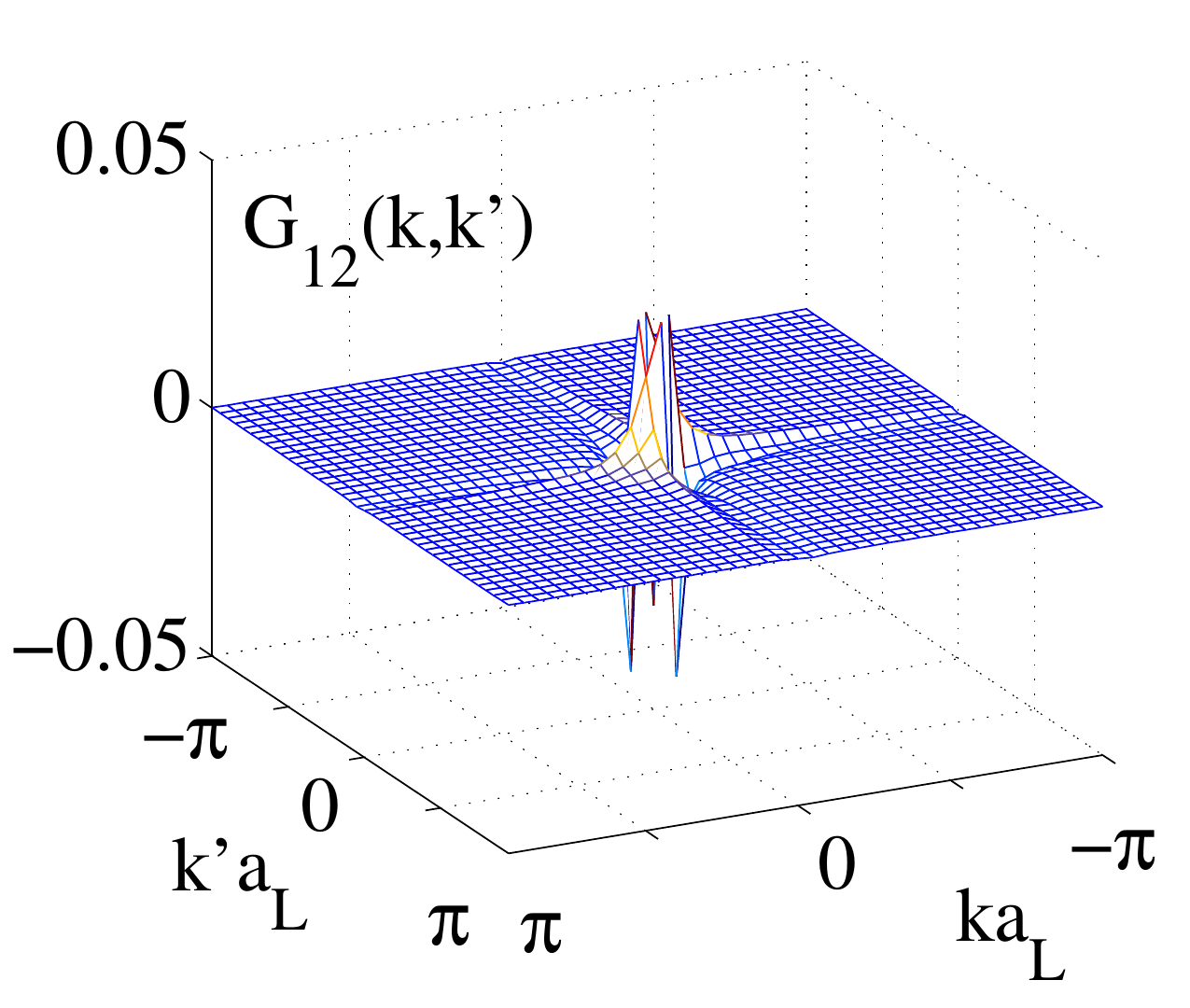}}\subfigure[PSF]{\includegraphics[width=4.5cm]{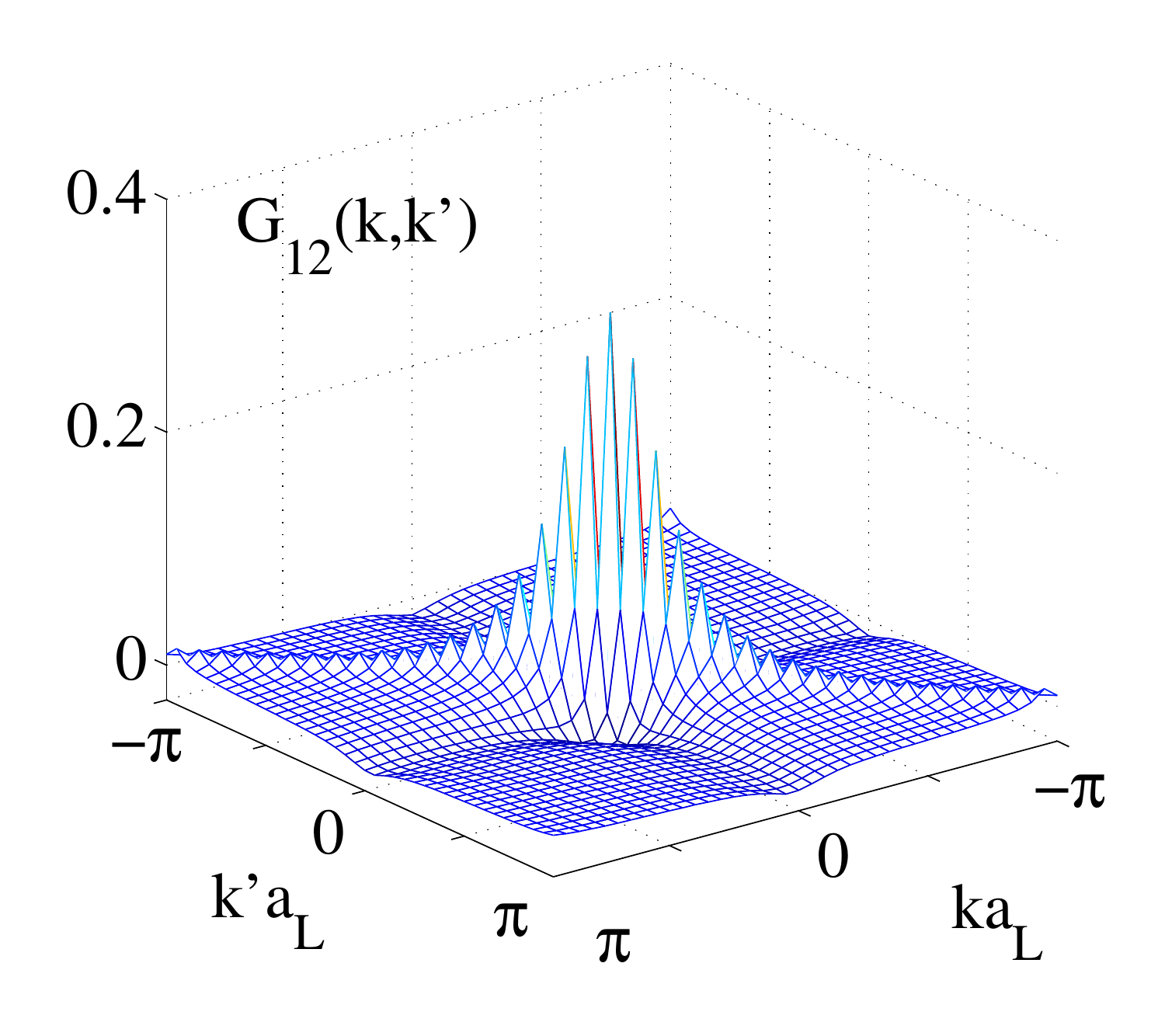}}\subfigure[CFSF]{\includegraphics[width=4.5cm]{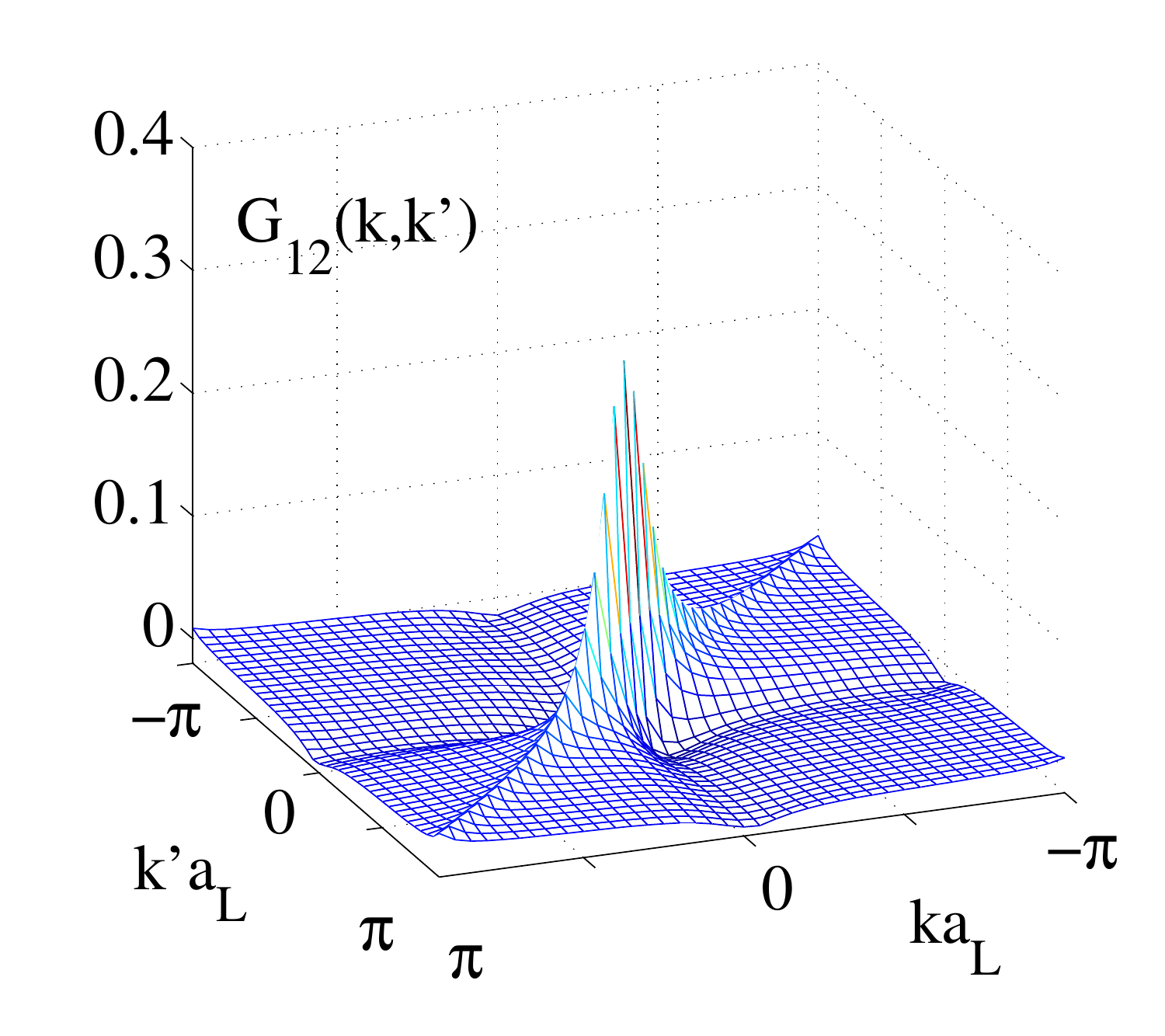}}

\caption{\label{fig:homo} Noise correlations for a homogeneous system of 40
lattice sites, calculated with the TEBD method. The frames (a) --
(d) correspond to the examples (a) -- (d) in Table \ref{tab:param}.
In (a), we show the noise correlations of a MI state. In the plot
of $\mathcal{G}_{11}(k,k')$, there is a strong correlation along
the direction $k=k'$, whereas the noise correlation function $\mathcal{G}_{12}(k,k')$
essentially vanishes. In (b), we show the noise correlations of a
SF state. Here, we can see the peak around $k=k'$ corresponding to
the $\delta$- function bunching peak predicted by LL theory (see
also Fig. \ref{fig:LL} (b)). For $\mathcal{G}_{12}$, we find negative
value at $k=k'=0$, which is different from the LL result (Fig. \ref{fig:LL}
(b)). Other structures predicted by LL theory can be seen in Fig.
\ref{fig:SF_NG}, where $\mathcal{G}_{12}$ and $\mathcal{G}_{11}$
are plotted in a non-linear color scale to magnify the structures
around $k=k'=0$. In (c) and (d), we show the noise correlations of
the PSF and CFSF state, respectively. In the PSF state (c), the inter-species
correlation $\mathcal{G}_{12}(k,k')$ has strong correlations along
$k=-k'$, a consequence of pairing (see also Fig. \ref{fig:LL} (c)).
In the CFSF state (d), the peak is formed along the direction $k=k'$,
an indication of anti-pairing in the CFSF state (see also Fig. \ref{fig:LL}).}

\end{figure*}

\begin{figure}
\includegraphics[width=4.5cm]{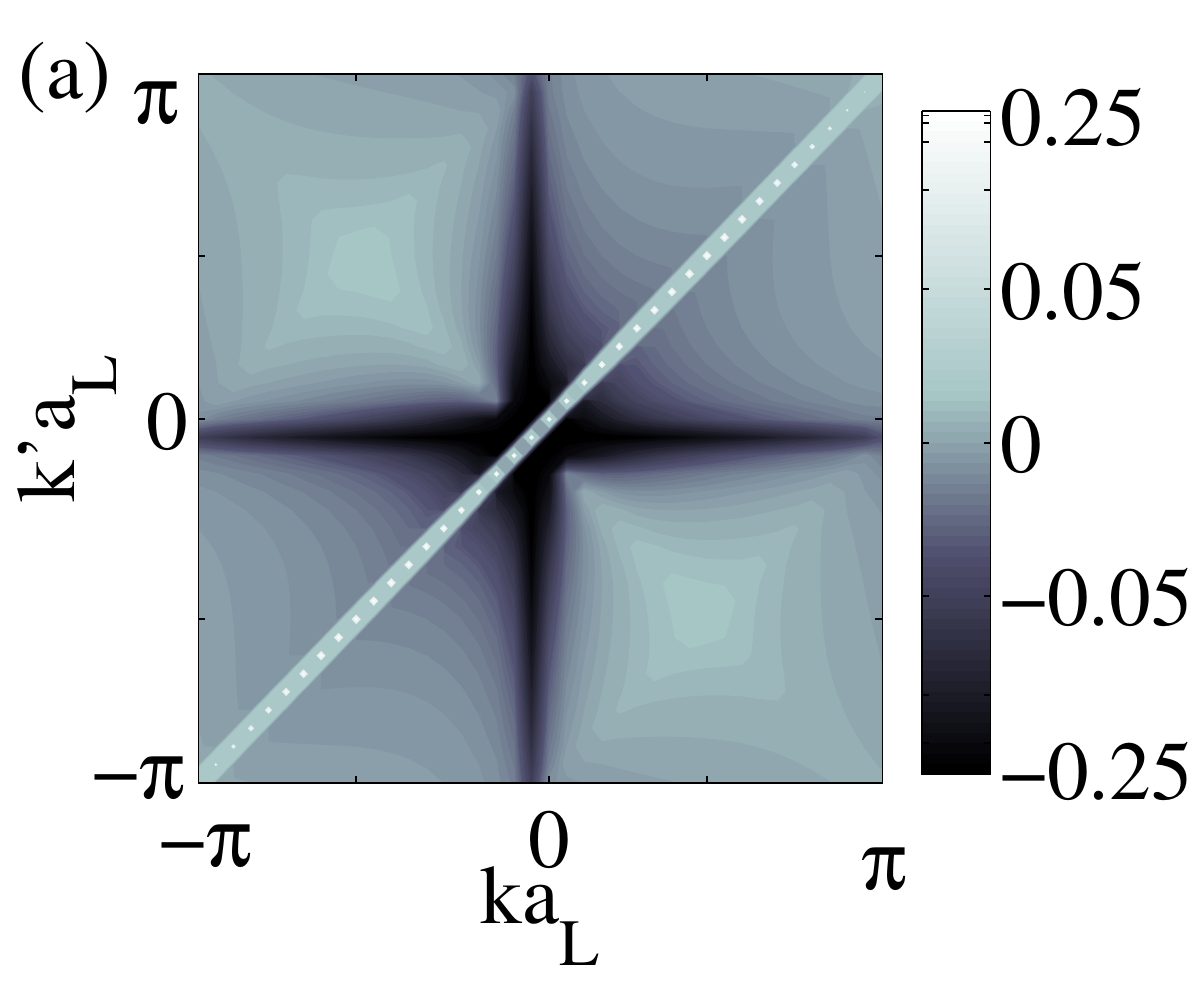}\includegraphics[width=4.5cm]{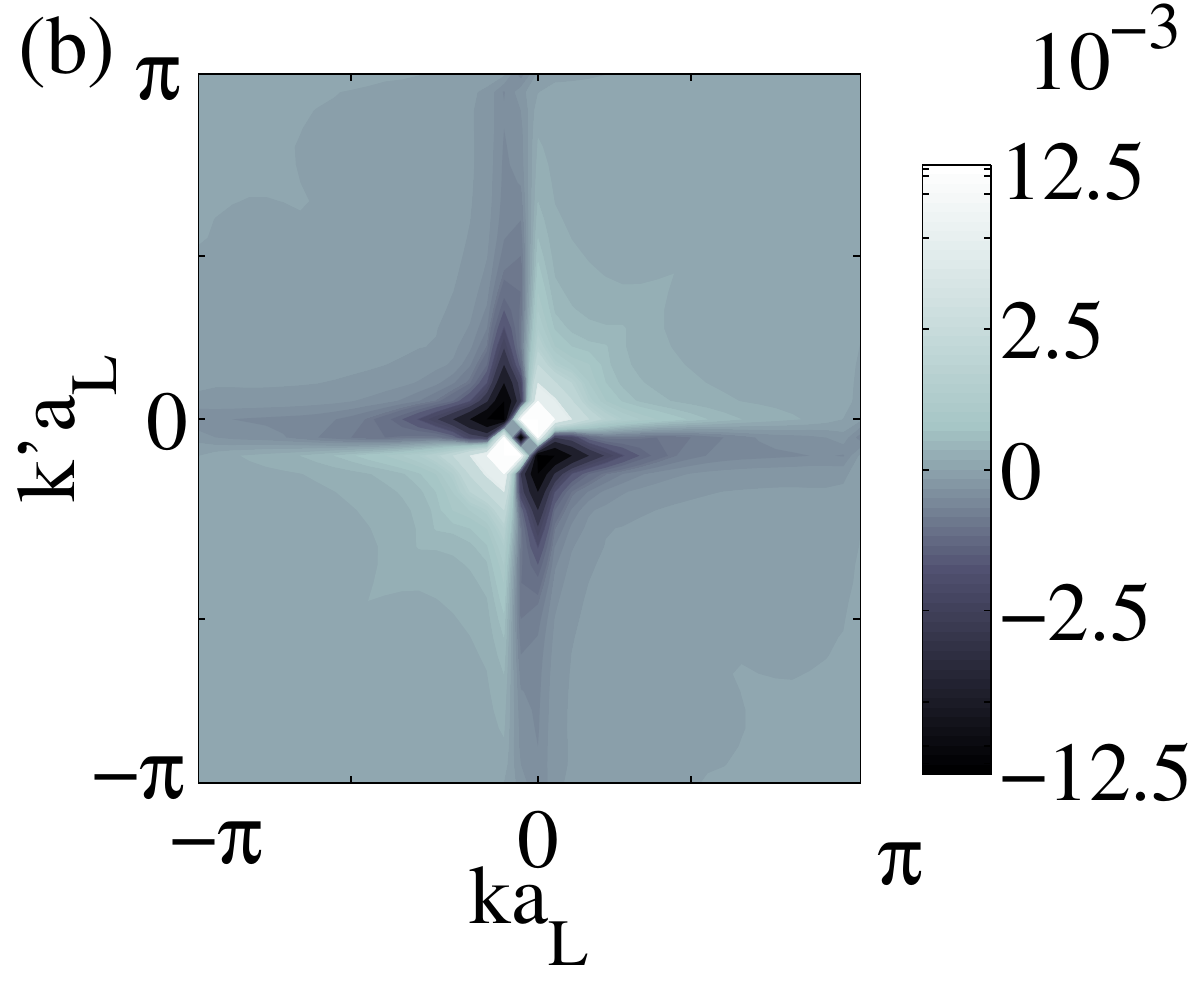}

\caption{\label{fig:SF_NG}Noise correlations, $\mathcal{G}_{11}(k,k')$ (a)
and $\mathcal{G}_{12}(k,k')$ (b), in the SF state of a homogeneous
system. The values of $\mathcal{G}_{11}(k,k')$ and $\mathcal{G}_{12}(k,k')$
are exactly the same as in Fig. \ref{fig:homo} (b). We create non-linear
gray scales by plotting $\tanh(10\mathcal{G}_{11})$ and $\tanh(200\mathcal{G}_{12})$
in linear scales. The labels of the color-bar reflects the values
of $\mathcal{G}_{11}(k,k')$ and $\mathcal{G}_{12}(k,k')$. The features
around $k=k'=0$ are magnified as a result of the non-linear scale.
In (a), we find the features predicted by LL calculations (\ref{fig:SF_LLG}
(a)). In addition, we can see a weak correlation at around $k=k'\pm2k_{F}$
, where $k_{F}=\nu\times\pi/a_{L}=0.5\pi/a_{L}$. This is where a
strong correlation (cusps) will develop if CDW order is present. This
feature can also been shown in LL calculations at around $k\thickapprox0$
and $k'\thickapprox2k_{F}$ (Eq. \ref{eq:noise_B2}). In (b), we find
that the structures along $k=k'$ is similar with the ones in LL calculations,
however, the structures along $k=-k'$ is negative, different from
the LL predictions (see also Fig. \ref{fig:SF_LLG}). The difference
may be understood as a result of different boundary conditions used
for the finite-size calculations: the numerical calculations use a
{}``hard-wall'' boundary condition, whereas the LL calculations
assume a periodic boundary condition.}

\end{figure}

\subsection{Homogeneous system}

In this section we discuss the numerical results for noise correlations
of a homogeneous system of 40 lattice sites, subject to the hard-wall
or {}``open'' boundary condition, in which the wave function is
required to vanish on the fictitious sites of index $0$ and $N+1$
implied by Eq. \ref{eq:hamiltonian}. We consider five parameter sets
listed in Table \ref{tab:param}, representing different regimes of
the phase diagram of 1D Bose mixtures.

\subsubsection*{Superfluid and Mott insulator}

For the Hamiltonian of Eq. \ref{eq:hamiltonian}, in the non-interacting
case, $U_{12}=0$, SF and MI are the only two possible orders. In
the interacting case, SF and MI orders are still encountered, when
the inter-species interaction is weak. For the Hamiltonian of Eq.
\ref{eq:hamiltonian} with $t\ll U$, the MI state exists for any
$|U_{12}|\lesssim U$, until the occurrence of collapse ($U_{12}\lesssim-U$)
or phase separation ($U_{12}\gtrsim U$). The SF state however exists
only when $|U_{12}|\ll U$. In either of SF or MI phases, the quasi-order
is formed in each individual species and the cross-species correlation
is weak.

For the MI state (Fig. \ref{fig:homo} (a)) we find that $\mathcal{G}_{11}(k',k)$
shows strong correlations along the direction $k'=k$, in agreement
with the LL theory result shown in Fig. \ref{fig:LL} (a). We also
find that the correlations along $k'=k$ are not uniform and that
the peak along $k'=k$ resembles a Lorentzian distribution in $k$
imposed upon a constant. This Lorentzian is due to the characteristic
scale of the correlation functions. This contribution was ignored
in the before-mentioned approximation in the LL calculation, but could
be included in a straightforward manner. The cross-species noise correlation,
$\mathcal{G}_{12}(k,k')$, on the other hand, is essentially zero,
indicating the absence of cross-species correlations in the MI state.
For the SF state (Fig. \ref{fig:homo} (b) and Fig. \ref{fig:SF_NG}),
we consider the case where there is weak repulsion between the two
species (Table \ref{tab:param} (2)). The Luttinger parameters are
$K_{A}=1.03$ and $K_{S}\simeq0.96$, which were extracted from the
correlation functions $R_{S}$ and $R_{A}$ by numerical fitting.
From the upper panel in \ref{fig:homo} (b), we see that $\mathcal{G}_{11}$
has the characteristic features of a quasi-condensate \cite{LM}:
the positive correlations along $k=-k'$, which indicate pairing;
the negative correlations between $k=0$ and finite $k'$, as well
as between $k'=0$ and finite $k$; and a $\delta$-function like
correlation along $k=k'$, corresponding to bosonic bunching. The
lower panel in Fig. \ref{fig:homo} (b), we see $\mathcal{G}_{12}$,
which shows similar features, except for the $\delta$-function along
$k=k'$, which is \char`\"{}softened\char`\"{} into a power-law divergence
and a slight negative value at $k=k'=0$. For a system of two non-interacting
superfluids, i.e. $U_{12}=0$, we have $K_{S}=K_{A}$, and $\mathcal{G}_{12}=0$. 

\begin{figure}
\includegraphics[width=4.3cm]{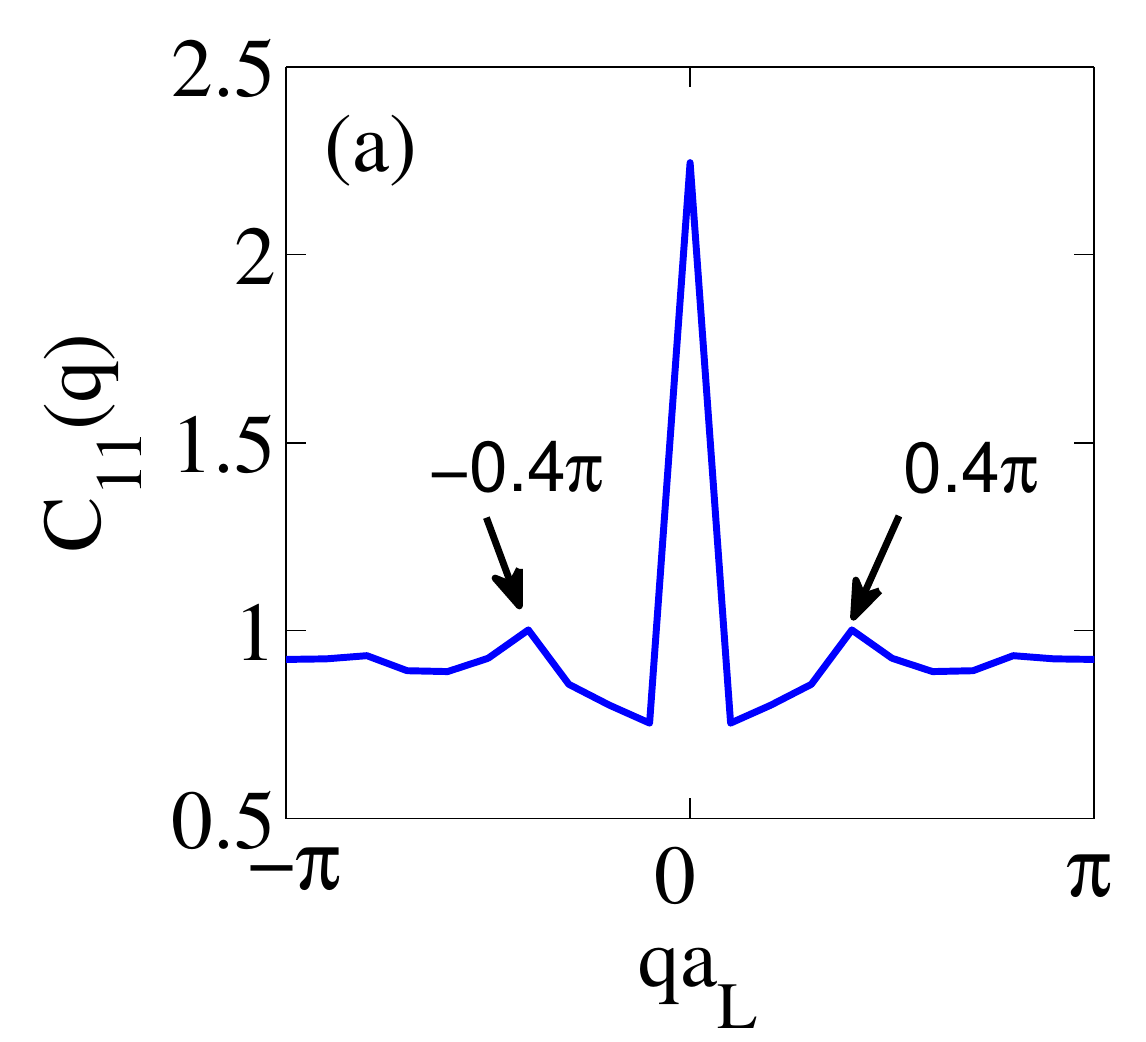}\includegraphics[width=4.3cm]{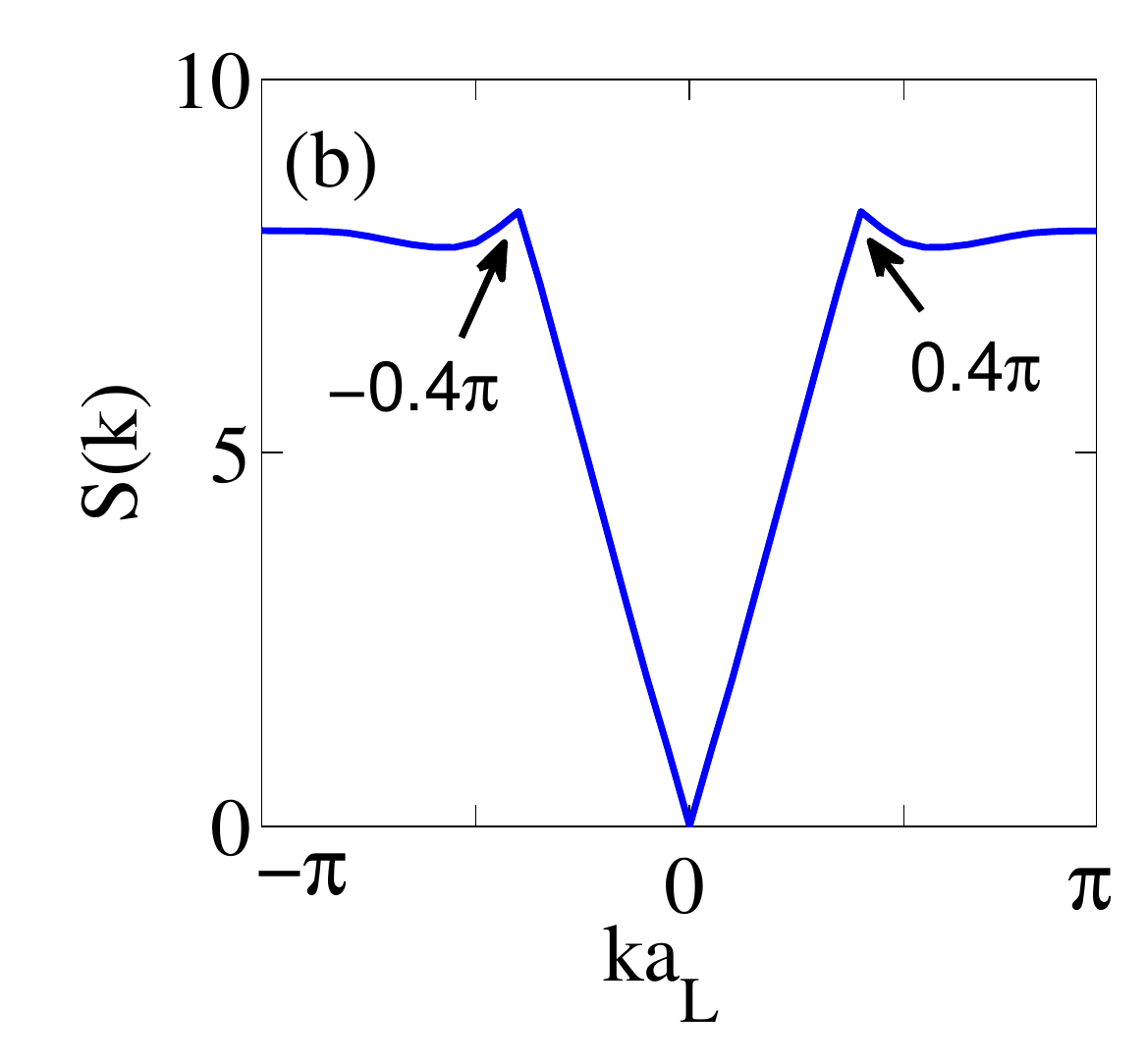}

\caption{\label{fig:CDW} Left : The correlation function $C_{11}(q)$, as
defined in Eq. \ref{eq:C_aa}; right: the structure factor $S(k)$
for a quasi-supersolid state. The parameters are given in example
(5) of Table. \ref{tab:param}. The Luttinger parameters are $K_{A}\approx1.4$
and $K_{S}\approx0.57$. The filling fraction is $\nu=0.2$, hence
the \char`\"{}Fermi wave vector\char`\"{} $k_{F}$ is $\pi\times0.2$.
At momentum $2k_{F}$, both quantities develop cusps, indicating the
presence of CDW order. }

\end{figure}

\subsubsection*{Paired superfluid and counter-flow superfluid }

We now discuss the noise correlations of the PSF and CFSF states.
The noise correlation $\mathcal{G}_{12}(k,k')$ is particularly important
for these two phases, because it can verify the existence of PSF and
CFSF orders. Unlike SF and MI states, PSF and CFSF states are characterized
by order parameters that contain both species and therefore cannot
be reflected in any single-species observables, such as the single-particle
Green's function, $G_{a}(x)$ or the single-particle momentum distribution
\cite{anzi}. The noise correlation function $\mathcal{G}_{12}(k,k')$
measures the correlations between the momentum occupancies of the
two species, and thus provides a direct probe of these orders. We
have shown in the previous section, that the peak along $k=-k'$ in
$\mathcal{G}_{12}(k,k')$ indicates the PSF order and that along $k=k'$
indicates the CFSF order. These features are verified in our numerical
calculation of $\mathcal{G}_{12}$ from the ground state.

In Fig. \ref{fig:homo} (c), we show the noise correlations in the
PSF state. The parameters are listed in (c) of Tab. \ref{tab:param}.
The existence of PSF order is - as usual - determined by the behavior
of the $R_{S}(x)$ and $R_{A}(x)$. $R_{A}(x)$ decays exponentially
and $R_{S}(x)$ algebraically with Luttinger parameter $K_{S}\simeq1.3$.
For the noise correlation function $\mathcal{G}_{12}(k,k')$, we find
that a peak is formed along $k=-k'$, which is a consequence of the
pairing correlations. In Fig. \ref{fig:homo} (d), we show our numerical
results for the CFSF example (d) in Table \ref{tab:param}. Based
on the behavior of the $R_{S}(x)$ and $R_{A}(x)$ we verify that
the system is in a CFSF state with $K_{A}\simeq1.2$, and an exponentially
decaying $R_{S}(x)$. For $\mathcal{G}_{12}$ we find that a peak
is formed along the diagonal direction, as a result of correlations
of anti-pairs ($b_{1}b_{2}^{\dagger}$). These findings are consistent
with the predictions of LL theory (see Fig. \ref{fig:LL}). We note
that $\mathcal{G}_{12}(k,k')$ is enhanced in magnitude in the PSF
and the CFSF phase compared to the MI and the SF phase, with a strongly
altered functional form.

\subsubsection*{Charge density wave }

In certain parameter regimes of the phase diagram, charge density
wave (CDW) order can coexist with each of the three superfluid orders,
SF, PSF and CFSF. In Sect. \ref{sec:Luttinger-Liquid-Calculation},
we use LL theory to show that CDW order can be reflected in the function
$\mathcal{G}_{11}$ and that the behavior of $\mathcal{G}_{11}$ around
$k=k'\pm2k_{F}$ resembles the structure factor $S(k)$. The reason
for the resemblance can be understood in a simple way, by recalling
the definition of the structure factor

\begin{equation}
S(k)=\frac{1}{N}\sum_{j_{1},j_{2}}e^{-ik(j_{1}-j_{2})}(\langle n_{j_{1}}n_{j_{2}}\rangle-\langle n_{j_{1}}\rangle\langle n_{j_{2}}\rangle).\label{eq:Sk}\end{equation}
 As mentioned in Sect. \ref{sec:The-Model} the density correlation
function is \char`\"{}contained\char`\"{} in the noise correlations
and the term ,\begin{eqnarray*}
 &  & \sum_{j_{1},j_{2}=1}^{N}\langle b_{1,j_{1}}^{\dagger}b_{1,j_{2}}b_{1,j_{2}}^{\dagger}b_{1,j_{1}}\rangle e^{i[k(j_{1}-j_{2})+k'(j_{2}-j_{1})]},\end{eqnarray*}
 is part of the full sum that needs to be taken for $\mathcal{G}_{11}$.
This term can also be written as a function of the density operator,
$n_{j}=b_{j}^{\dagger}b_{j}$ , as \begin{eqnarray*}
 &  & \sum_{j_{1},j_{2}=1}^{N}\langle n_{j_{1}}n_{j_{2}}\rangle e^{-i(k'-k)(j_{1}-j_{2})}+\delta(k-k')\langle n_{j_{1}}\rangle e^{-i(k'-k)j_{1}}\end{eqnarray*}
 This shows that $\mathcal{G}_{11}$ and $S(k)$ (Eq. \ref{eq:Sk})
have the same Fourier transform of the density correlation function.
If $S(k)$ develops cusps at $\pm2k_{F}$, where $k_{F}=\pi\nu$ \cite{anzi},
when CDW order is present, we expect $\mathcal{G}_{11}$ to have similar
cusps at $k=k'\pm2k_{F}$. In Fig. \ref{fig:CDW}, we show one example
of a quasi-supersolid (SS) state \cite{supersolid_LM}, where CDW
order coexists with SF order. The parameters are listed in (e) of
Table \ref{tab:param}. In the plot, the correlation function $C_{11}(q)$,
an integration of $\mathcal{G}_{11}(k,k')$ along the direction $k=k'$
(Eq. \ref{eq:C_aa}), is compared with the structure factor $S(k)$
of the same state. In both functions, we can see cusps appearing at
$\pm2k_{F}$.

\begin{figure}
\subfigure[MI]{\includegraphics[width=4cm,height=4cm]{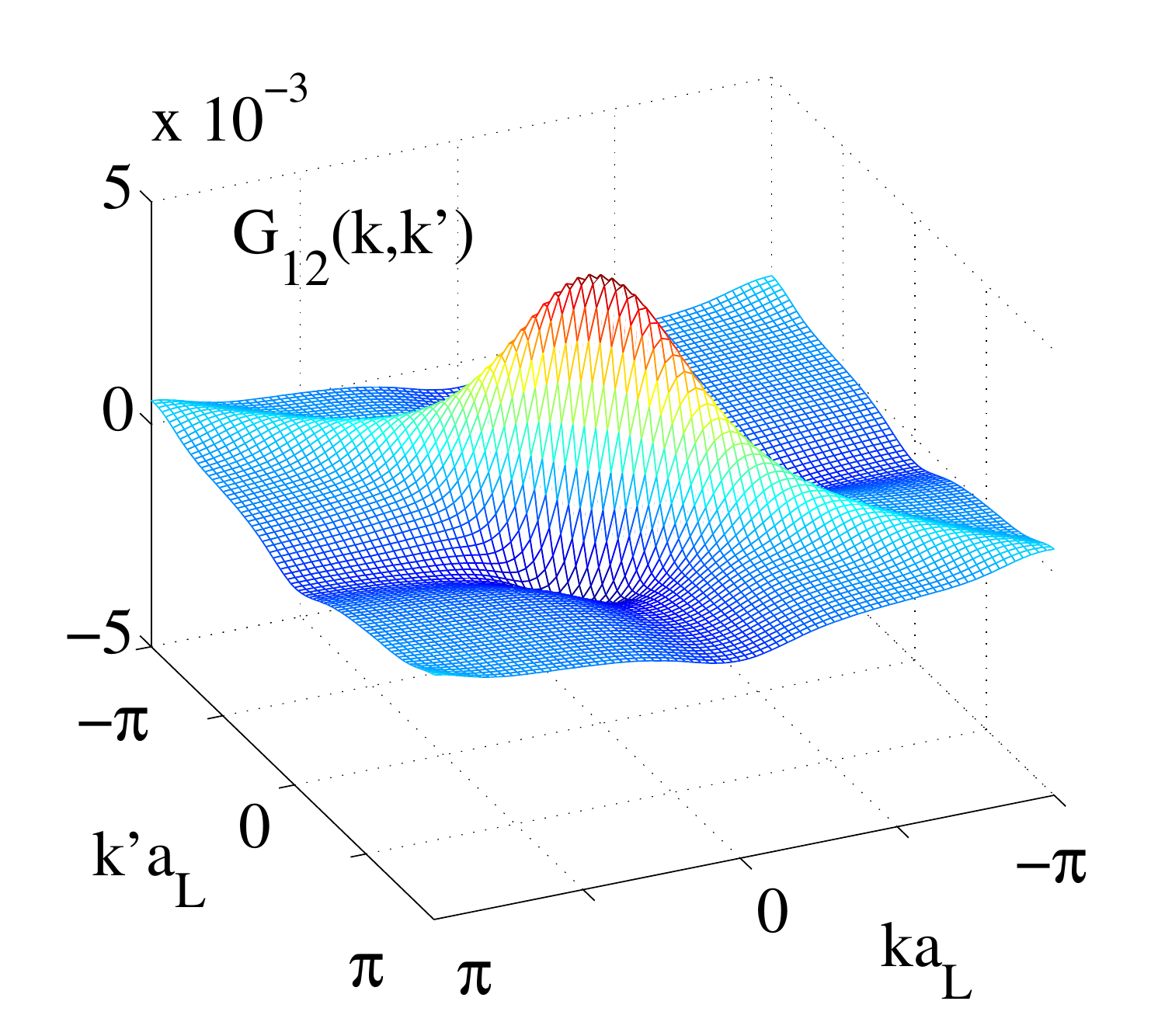}}\subfigure[SF]{\includegraphics[width=4cm,height=4cm]{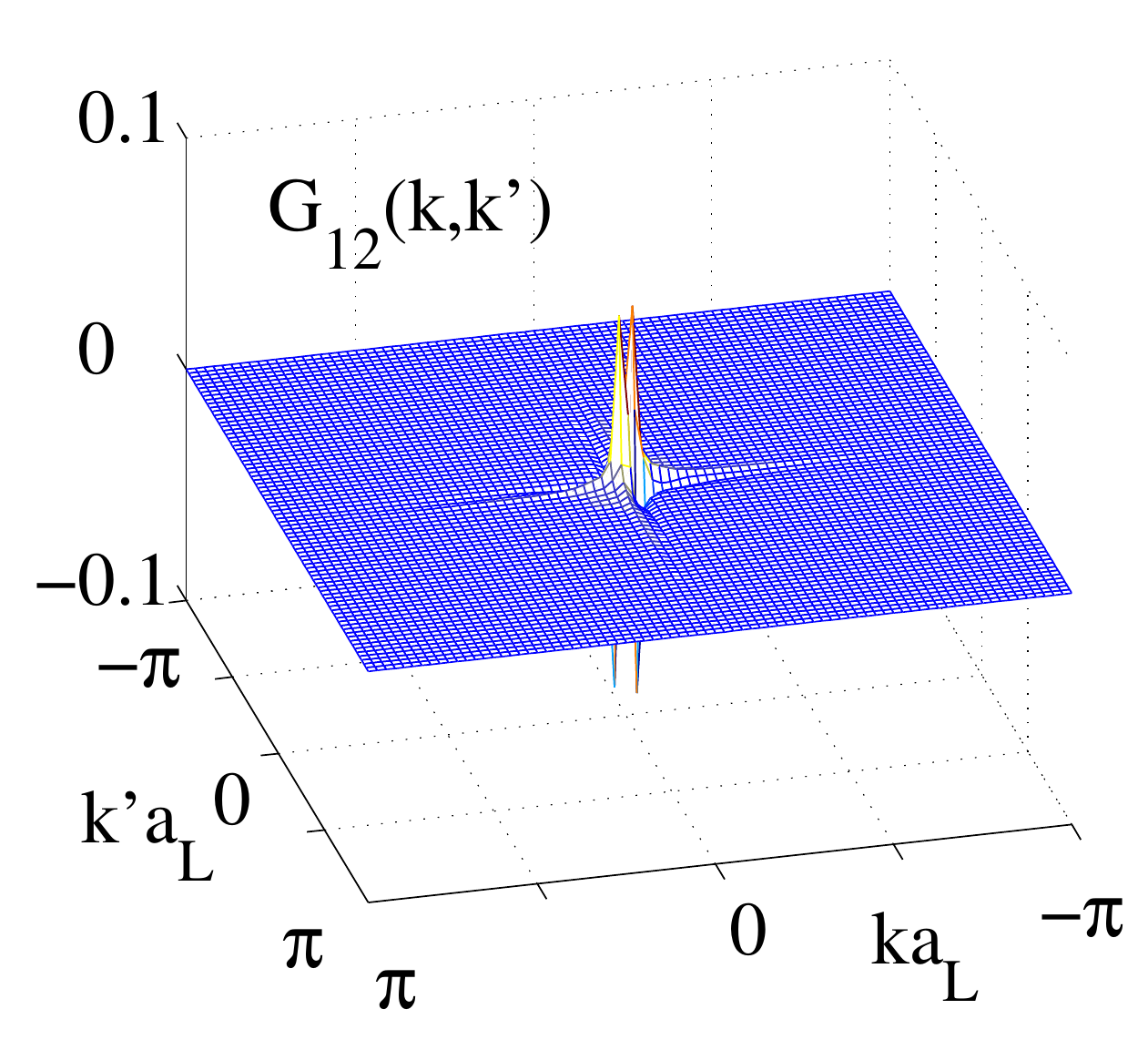}}

\subfigure[PSF]{\includegraphics[width=4cm,height=4cm]{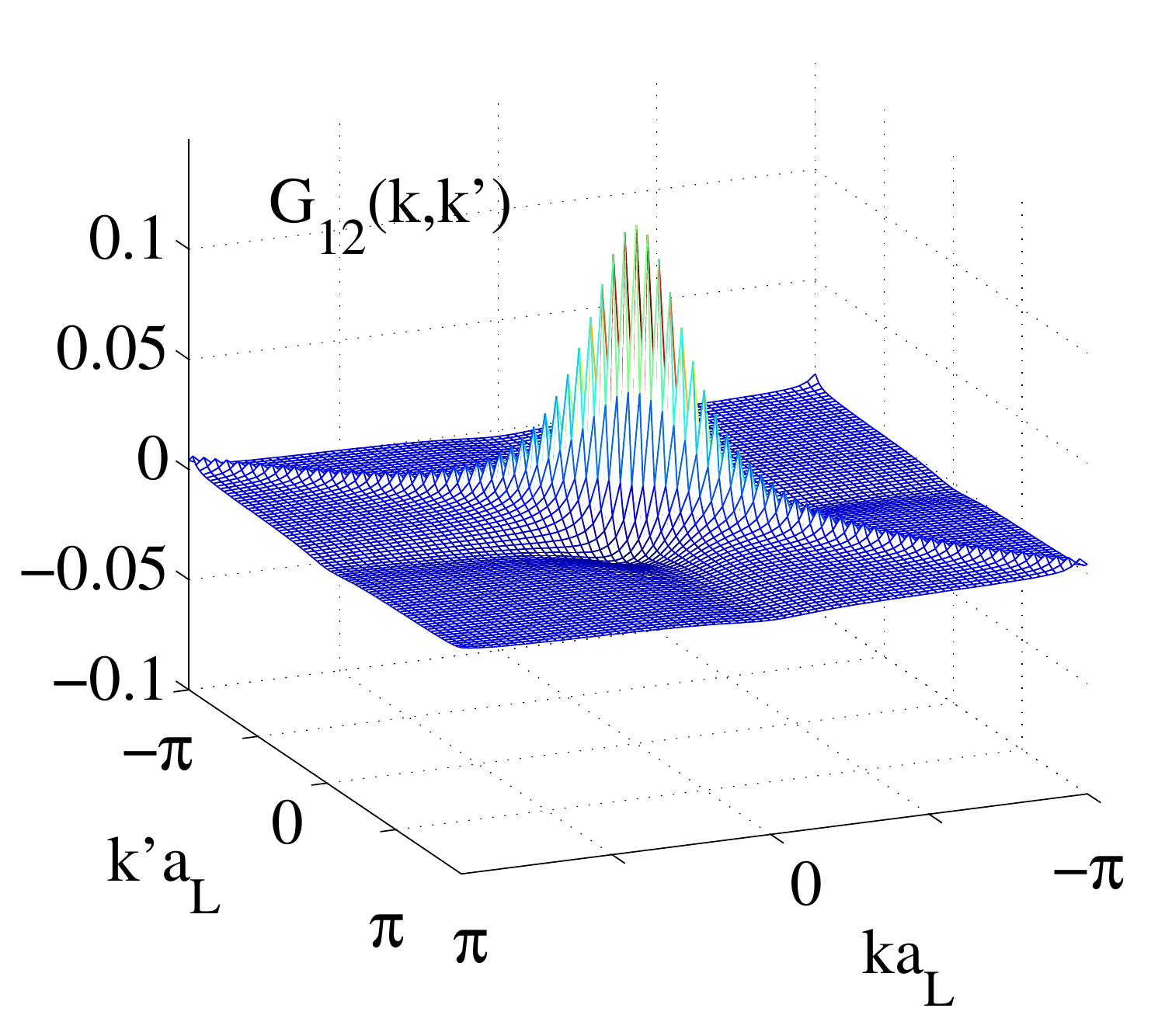}}\subfigure[CFSF]{\includegraphics[width=4cm,height=4cm]{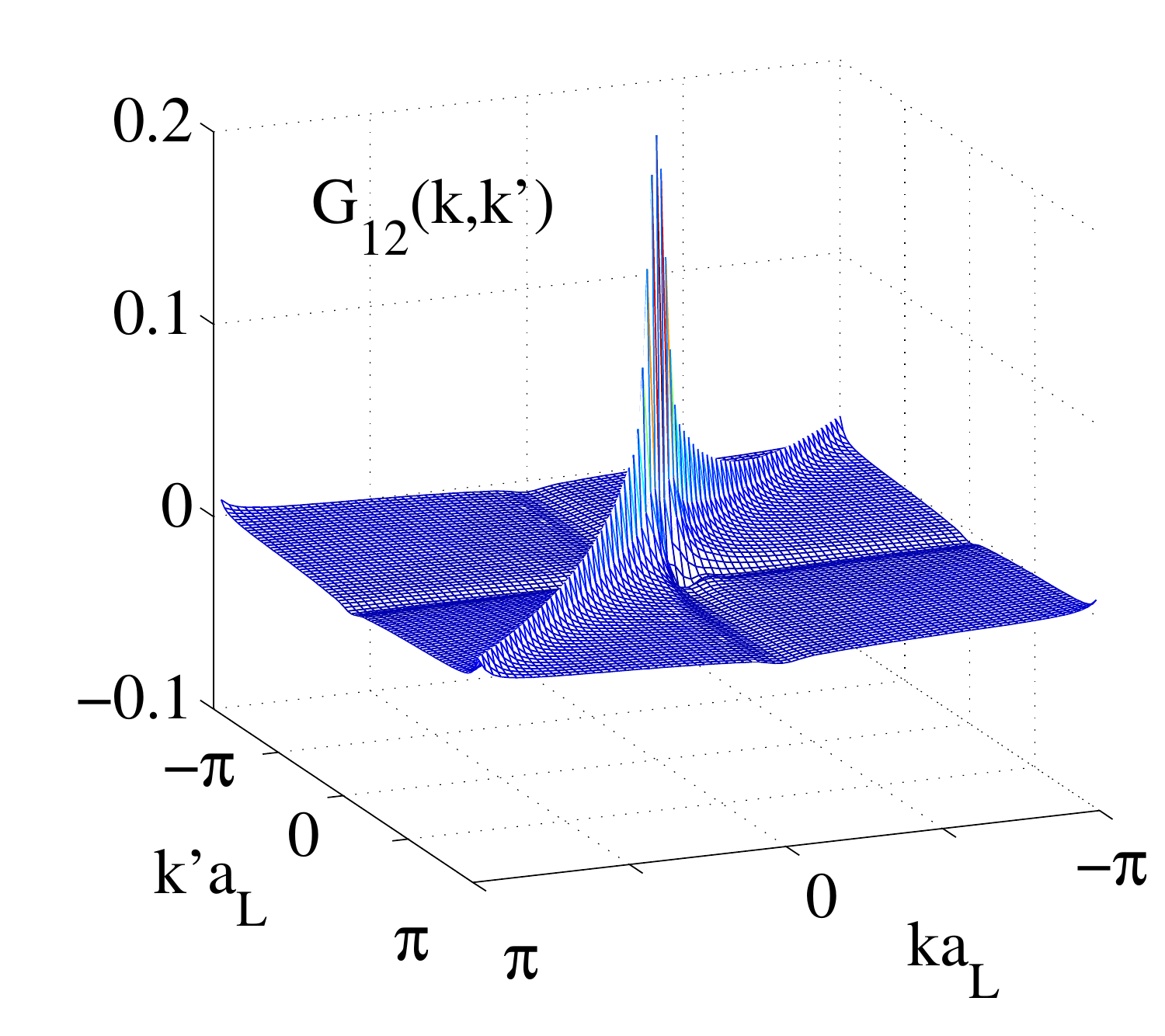}}

\caption{\label{fig:trap1}Noise correlations in the trapped system. The system
size is 80 sites and $t/U=0.02$. In (a), the system is in the SF
state. The particle number of each species is 30, the trap frequency
$8\times10^{-5}U$ and $U_{12}/U=0.01$. In (b), the particle number
of each species is 40, the trap frequency $1\times10^{-4}U$ and $U_{12}/U=-0.11$.
The system has both MI and PSF orders. The MI state forms a plateau
at unit-filling at the center of the trap and the PSF is formed at
the edge. The PSF state at the edge causes the small peak along the
$k=-k'$ direction, similar to the one in (c). However, this peak
is at a much smaller amplitude than the one shown in (c), where the
whole system is a PSF state. In (c), the particle number of each species
is 20, the trap frequency $1\times10^{-5}U$ and $U_{12}/U=-0.11$.
The whole system is in the PSF state. A strong pairing correlation
is formed along $k=-k'$ direction. In (d), the particle number of
each species is 30, the trap frequency is $8\times10^{-5}U$ and $U_{12}/U=0.2$.
The system has both CFSF and SF order. The CFSF order forms a plateau
at half-filling at the center of the trap and the SF state towards
the edges of the trap. The CFSF order causes a strong anti-pairing
(particle-hole) correlation along $k=k'$direction. At the same time,
the SF order adds to the \char`\"{}dips\char`\"{} along $k=0$ and
$k'=0$. }

\end{figure}

\subsection{Noise correlations in the trapped system}

We now discuss how the different types of order are affected by the
presence of a trapping potential. To simulate the effect of a trap,
we add a harmonic potential, $\Omega(j-j_{c})^{2}(n_{1,j}+n_{2,j})$
to the Hubbard Hamiltonian in Eq.~\ref{eq:hamiltonian}, where $j$
is the site index and $j_{c}$ is the index at the center of the system.
We then use the TEBD method to calculate the ground state. We also
increase the system size to 80 lattice sites, and choose the total
number of particles and the trap frequency to ensure that the boundary
effect is negligible.

\begin{figure}
\includegraphics[width=4.5cm]{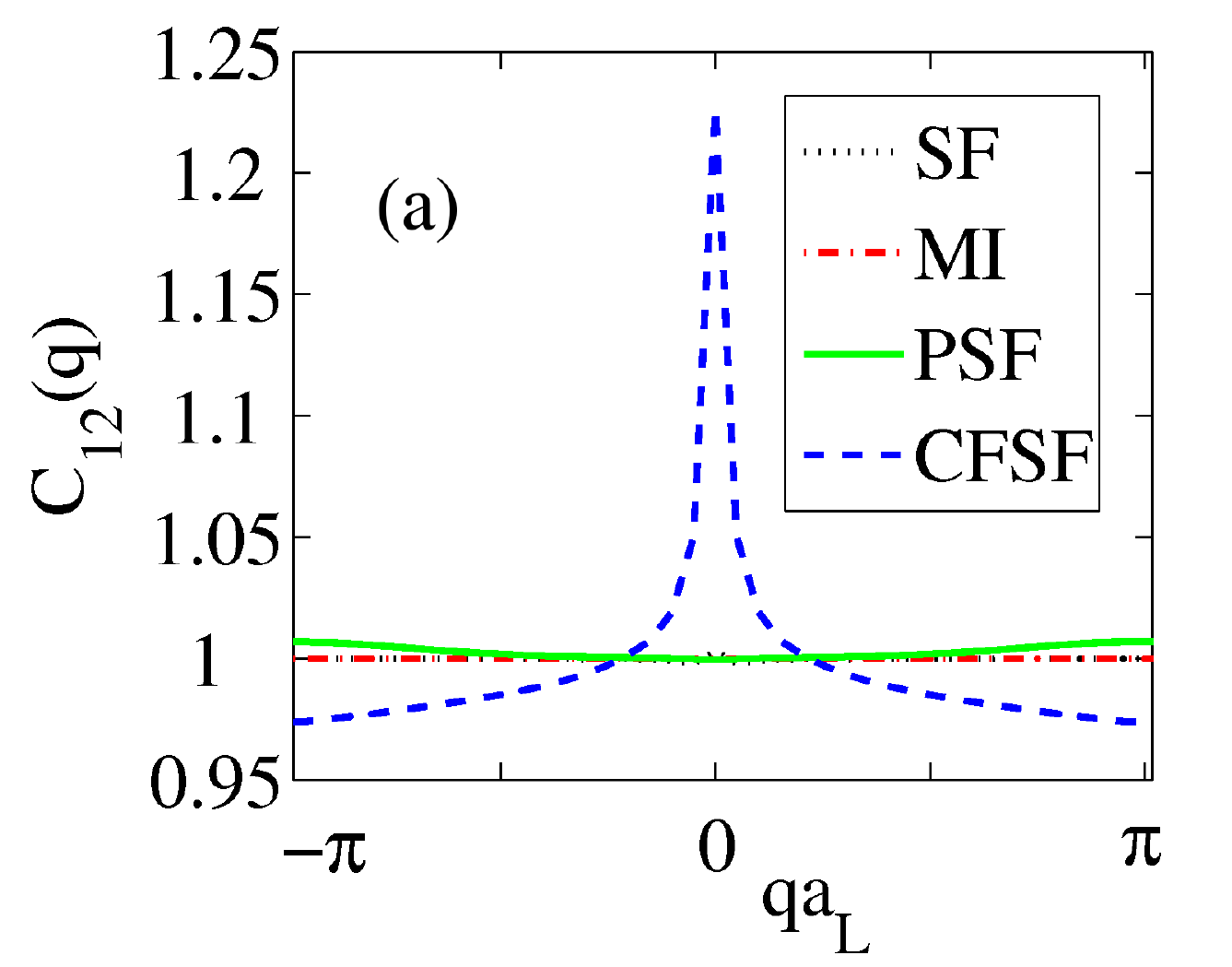}\includegraphics[width=4.5cm]{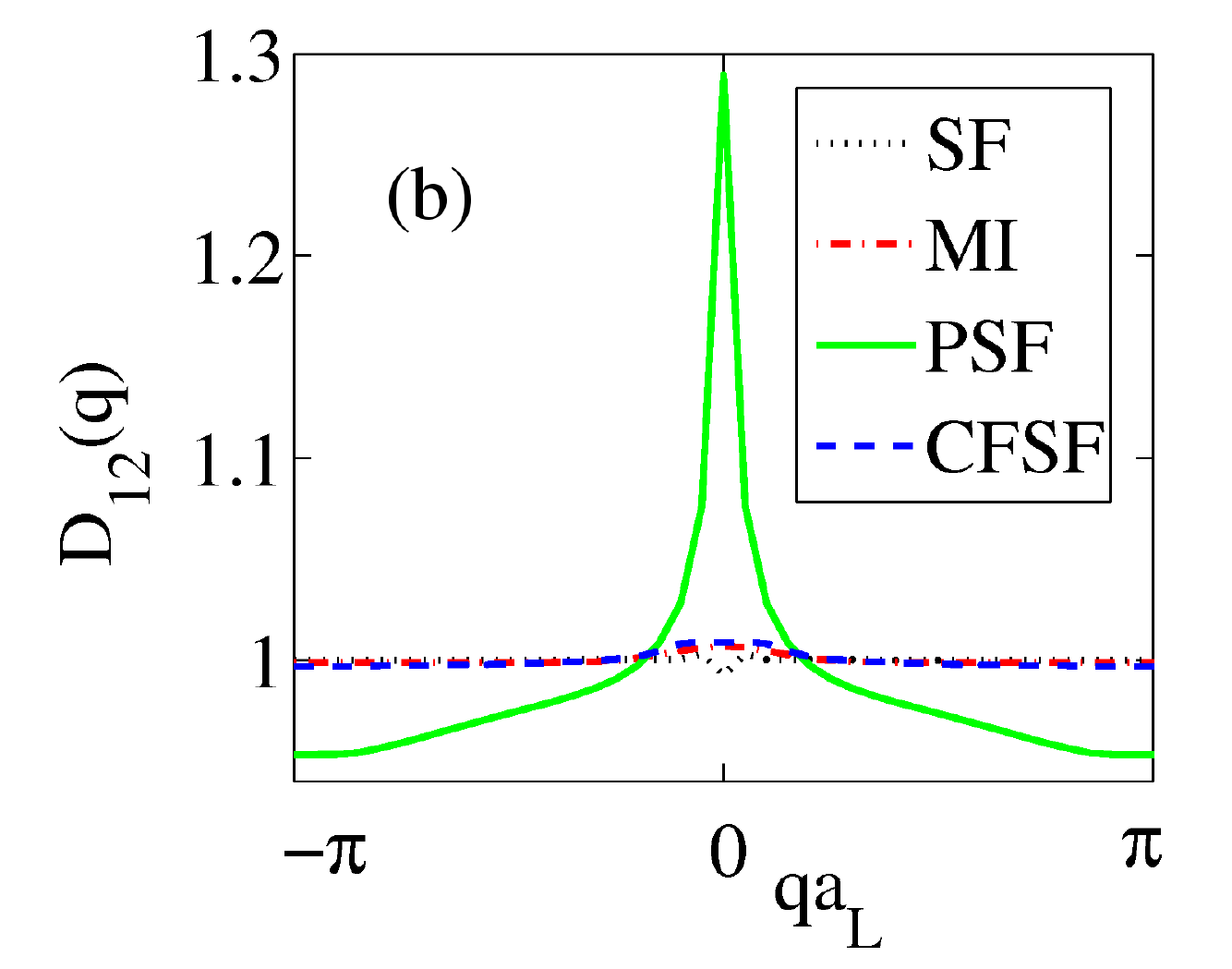}

\caption{\label{fig:trap_C}Correlations $C_{12}(q)$ and $D_{12}(q)$ for
the states that are described in Fig. \ref{fig:trap_C}. In (a), we
show the behavior of $C_{12}(q)$ (Eq. \ref{eq:C_aa}) in SF, MI,
PSF and CFSF states. The strong anti-pairing (particle-hole) correlations
in the CFSF state gives a strong signal around $q=0$ in $C_{12}(q)$.
This strong signal is also unique to the CFSF state and therefore
can be used to detect to the CFSF order. In (b), we show the behavior
of $D_{12}(q)$ (Eq. \ref{eq:D_aa}) in SF, MI, PSF and CFSF states.
The strong pairing correlation in the PSF state is the reason for
the high peak around $q=0$ in $C_{12}(q).$ This suggests measuring
$C_{12}(q)$ is a good way of detecting the PSF order.}

\end{figure}

One interesting feature of a trapped system is that different orders
can coexist in the trap. A well-known example is the MI plateau at
the center of the trap surrounded by a SF at the edge \cite{folling_shell}.
For repulsive inter-species interaction, we find coexistence of a
CFSF plateau with a SF at its edge and a MI plateau with PSF at the
edges for attractive inter-species interactions \cite{anzi}. Despite
the potential complication of coexistence of orders, we find clear
signals for the pairing correlations of the PSF phase and the anti-pairing
correlations of the CFSF phase.

\begin{figure}
\includegraphics[width=4.3cm]{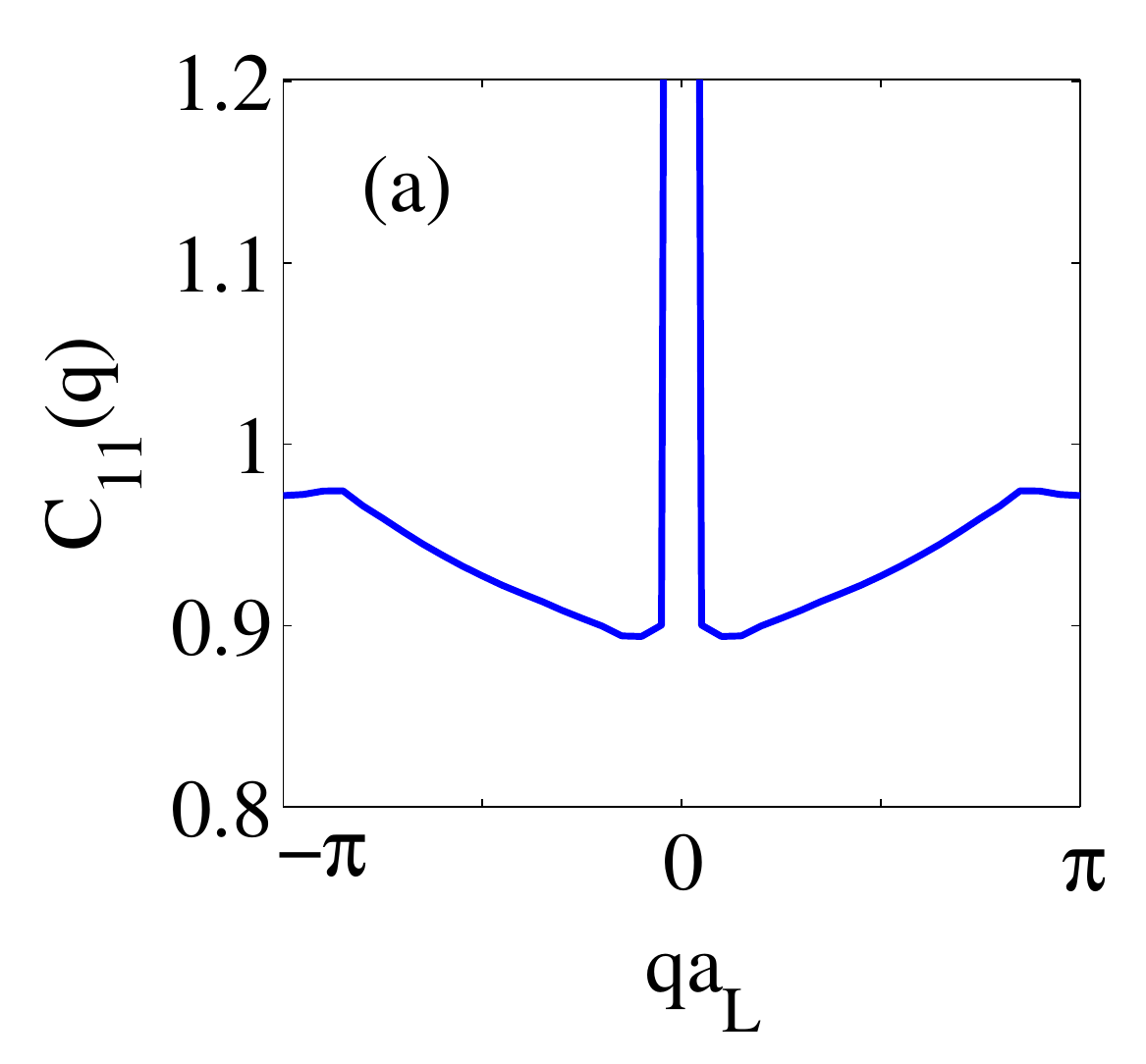}\includegraphics[width=4.3cm]{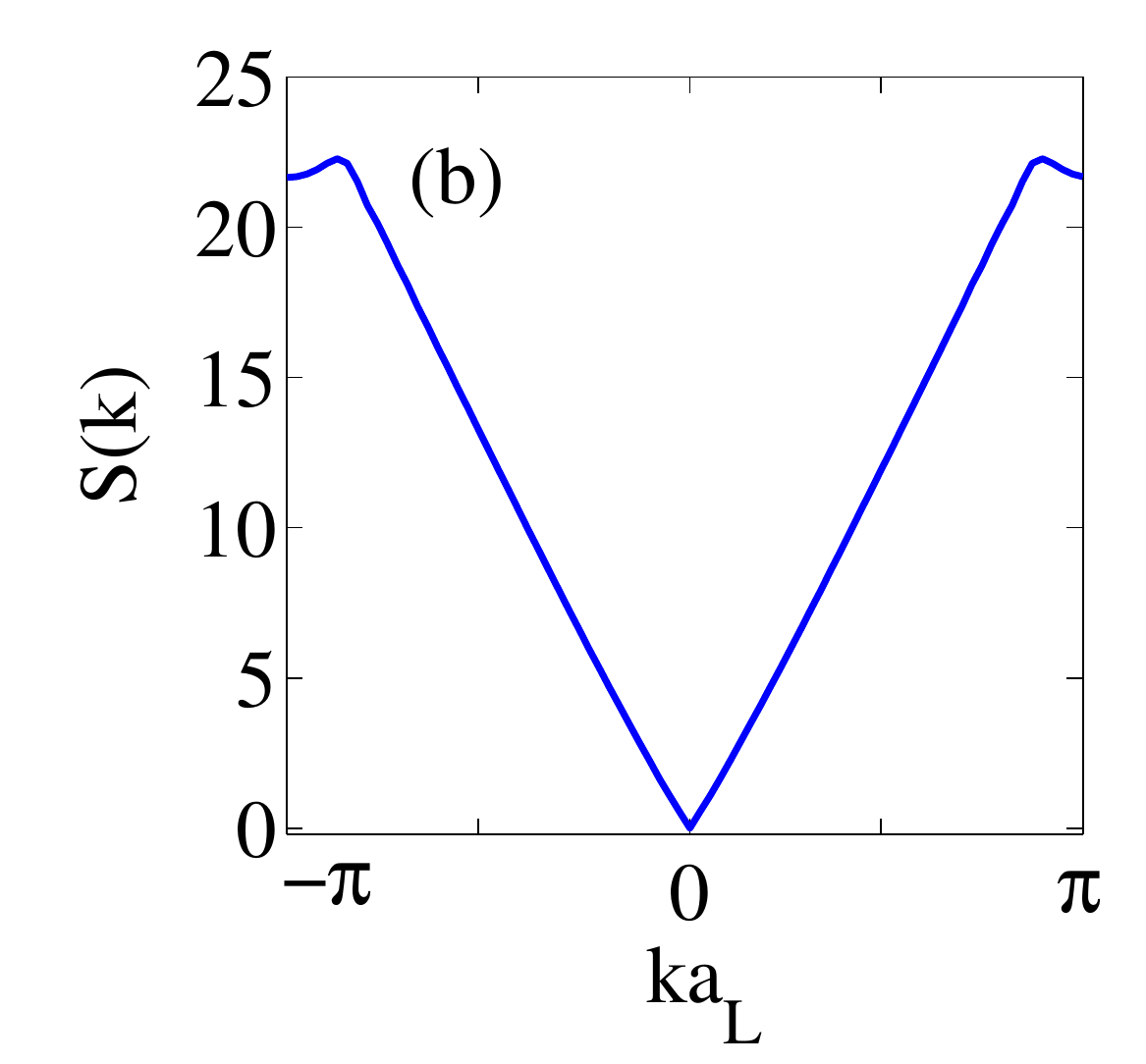}

\caption{\label{fig:trap_CDW}Noise correlation $C_{11}(q)$ and structure
factor $S(k)$ in a PSF/CDW state. The system size is 80 sites and
there are 20 particles of each species ($\nu=0.25$). The trap frequency
is $\Omega=10^{-5}U$ , the hopping $t=0.02U$ and the inter-species
interaction is $U_{12}=-0.11U$. The density at the center of the
trap is roughly $0.45$ per site and the cusps are developed around
$\pm0.9\pi$. The inhomogeneity of a trapped system means that the
{}``Fermi wave vector'' $k_{F}$ is no longer $\pi\nu$, where $\nu$
is the average filling of the system. Instead, $k_{F}$ can be evaluated
as $\pi n_{center}$, where $n_{center}$ is the density at the center
of the trap, }

\end{figure}

In Fig. \ref{fig:trap1}, we show the behavior of $\mathcal{G}_{12}(k,k')$
in four different cases, where the orders at the center of the trap
are SF, MI, PSF and CFSF respectively. We find that the general behavior
of the noise correlation in a trap is very similar to its homogeneous
counterpart. In Fig. \ref{fig:trap1} (c) and (d), $\mathcal{G}_{12}$
shows clearly the feature of pairing correlations in the PSF state
and the anti-pairing correlations in the CFSF state. In addition,
we see some minor features attributed to the coexisting orders. In
the case of CFSF in a trapped system, we can see the \char`\"{}dip\char`\"{}
along $k=0$ and $k'=0$ because of the coexistence with the SF order.
On the other hand, in the case of a MI in a trap, we can see pairing
correlations as a result of the residual PSF state at the edges. This
pairing signal is much smaller than when the whole system is in the
PSF state.

To show that the peaks along $k=k'$ and $k=-k'$ in $\mathcal{G}_{12}$
are detectable in experiments, we also calculate $C_{12}(q)$ (Eq.
\ref{eq:C_aa}) and $D_{12}(q)$ (Eq. \ref{eq:D_aa}) for the four
states. In the correlation $C_{12}(q)$ (Fig. \ref{fig:trap_C} (a)),
a high peak at $q=0$ \emph{only} appears in the case of the CFSF
state. This peak corresponds to the peak in $\mathcal{G}_{12}$ along
$k=k'$ in the CFSF state and is a reflection of the anti-pair correlation
in the CFSF state. Similarly, in $D_{12}(q)$, the high peak at $q=0$
\emph{only} appears in the PSF state, as a result of the strong pairing
correlations in the PSF state. A similar measurement has been performed
for fermionic mixtures to detect the pairing of fermions \cite{greiner}.

In addition to the PSF and CFSF order, we also look for the signal
of CDW order in $\mathcal{G}_{11}(k,k')$ in the trapped system. In
a trapped system, the CDW order is more difficult to establish especially
in the PSF and SF states, because the varying local density makes
the \char`\"{}Fermi wave vector\char`\"{} $\pi n$ a spatially varying
quantity. However, we can still see weakened cusps forming at the
momentum roughly corresponding to $2\pi n_{center}$, where $n_{center}$
is the density at the center of the trap. This may indicate that in
the trapped system, the CDW order in PSF and SF states has a wave
vector corresponding to the density at the center of the trap. For
the CFSF state, because the system has a plateau at half filling,
the wave vector $2k_{F}$ is $\pi/a_{L}$. Compared to the homogeneous
case, this feature is slightly diminished due to the effect of the
coexisting SF state in the trapped system. In Fig. \ref{fig:trap_CDW},
we show one case where CDW order coexists with PSF order in a trap.
The system size is 80 sites and there are 20 particles of each species.
The trap frequency $\Omega=10^{-5}U$ , $t=0.02U$ and $U_{12}=-0.11U$.
The density at the center of the trap is roughly $0.45$ per site.
The cusps are developed around $\pm0.9\pi$, which is roughly $2\pi n_{center}$.

\subsection{Determination of Luttinger parameters from experimental data\label{sub:Determination}}

Another important question is whether we can use the noise correlation
$\mathcal{G}_{12}$ to measure the Luttinger parameters, $K_{S}$
and $K_{A}$, in the PSF and CFSF regimes. The LL calculation shows
that as the system size approaches infinity, the noise correlation
$\mathcal{G}_{12}$ approaches a power law decay with the power $-1/K_{S}$
in the PSF regime and with $-1/K_{A}$ in the CFSF regime (see Eqs.
\ref{eq:G_PSF} and \ref{eq:G_CFSF}). In our numerical results for
$C_{12}(q)$ and $D_{12}(q)$, we indeed find that the decay from
the peak at $q=0$ satisfies the algebraic decay. To find out the
power of the algebraic decay, we fit the function $C_{12}(q)$ in
the PSF regime and $D_{12}(q)$ in the CFSF regime with the fitting
function,

\begin{equation}
F(q)=A|\mathrm{sin}(2q)|^{-1/K}+B,\label{eq:Fitting1}\end{equation}
 where $B$ is the minimum value of $C_{12}(q)$ or $D_{12}(q)$ and
$A$ and $K$ are the fitting parameters. In the PSF case ($U_{12}/U=-0.11$),
we find that $K$ is $1.3\pm0.1$. This is indeed very close to the
value of $K_{S}$, which is estimated at $1.4\pm0.1$ obtained by
the algebraic fit of $R_{S}$. In the CFSF case ($U_{12}/U=0.2$),
we find that $K$ is roughly $1.48\pm0.1$, while the value of $K_{A}$
extracted from the algebraic fit of $R_{A}$ is also at $1.48\pm0.12$.
Because of the singularity at $q=0$, a reasonable values of $K$
can be obtained by a simple algebraic decay function, $Aq^{-1/K}+B$,
around small $q$. This shows that even in a trapped system, one can
still assume a algebraical relationship predicted in the LL theory
(Eqs. \ref{eq:G_PSF} and \ref{eq:G_CFSF}) and estimate the values
of the Luttinger parameters by studying the power of the decay from
the peak at $q=0$.

\section{\label{sec:Conclusion}Conclusions}

We have studied the behavior of noise correlations for a binary bosonic
mixture in optical lattices. We consider different regions of the
phase diagram and we show that the noise correlations have different
signatures for different phases. In particular, we discuss the measurement
of the noise correlations as a means for detection of the paired superfluid
(PSF) and counter-flow superfluid (CFSF) order. Our study of a harmonically
trapped system shows that the inhomogeneity modifies the noise correlation,
due to the coexistence of different orders within the trap. These
modifications can be understood in terms of the results for the homogeneous
system. What we find very encouraging is that even with the presence
of a trap, the noise correlations still have distinctive features
for each order, and the peak structure of the noise correlation $\mathcal{G}_{12}$
in the PSF/CFSF regime still obeys the algebraic decay relationship
predicted by the LL theory. This means that one can use the noise
correlation to estimate the Luttinger parameters in these two regimes.
All these results would be useful for experiments aimed at detecting
the pairing and anti-pairing orders that can exist in ultracold atom
systems.

\acknowledgments We thank I. Danshita for useful discussions. This
work was supported by the National Science Foundation under Physics
Frontiers Center Grant PHY-0822671. L.M. acknowledges support from
a NRC/NIST fellowship.

\end{document}